\documentclass[twocolumn]{aastex6}

\usepackage{newtxtext,newtxmath}

\usepackage[T1]{fontenc}
\usepackage{ae,aecompl}

\usepackage{graphicx}	
\usepackage{amsmath}	
\usepackage{amssymb}	
\usepackage{bm}
\usepackage{subfig}
\usepackage{mwe}

\usepackage{todonotes}
\newlength{\tempheight}
\newlength{\tempwidth}

\newcommand{\rowname}[1]
{\rotatebox{90}{\makebox[\tempheight][c]{\textbf{#1}}}}

\newcommand{\columnname}[1]
{\makebox[\tempwidth][c]{\textbf{#1}}}

\shorttitle{Magnetised NS head on collision}
\shortauthors{Nathanail Antonios}

\begin{document}
\title{A Toy Model for the Electromagnetic Output of Neutron-star Merger Prompt Collapse 
to a Black Hole: Magnetized Neutron-star Collisions}
\author{Antonios Nathanail}\thanks{E-mail: nathanail@th.physik.uni-frankfurt.de}
\affil{Institut f\"{u}r Theoretische Physik, Max-von-Laue-Strasse 1, D-60438 Frankfurt, Germany}




\label{firstpage}

\begin{abstract}
We present a systematic study of magnetised neutron 
star head on collisions. We investigate 
the resulting magnetic field geometries as the two neutron stars 
merge. Furthermore, we analyze the luminosity produced in these collisions
and monitor the evolution of the magnetic fields from the time of merger
until the subsequent production of a black hole. At the time of black 
hole formation the luminosity peaks and rings-down following the decay 
of the electromagnetic fields. A comparison is 
presented for three different cases, one where the initial magnetic field in both 
neutron stars is aligned, one where they are anti-aligned and also
one case where they initially have unequal magnetic field strength. 
We identify regions and set limits so that pair creation and magnetic reconnection
would occur in this scenario, and further discuss limits and differences in the 
radiated energy. 
This study should be regarded as a toy model of the case where the remnant,
 of a binary neutron star merger, undergoes a prompt collapse to a 
 black hole with a negligible surrounding disk.
We note that the generated electromagnetic pulses resembles 
the fast radio bursts phenomenology. We consider implications on the
high mass mergers leading to a fast prompt collapse to a black hole 
and the expected flux to be observed at a distance similar to the 
binary neutron star gravitational wave detection GW190425. 
\end{abstract}




\section{Introduction}
\label{sec:intro}

The coincident detection of gravitational waves (GW) from binary neutron star 
mergers and the detection of a short gamma-ray burst 
(sGRB) together with the subsequent detection of afterglows across the
electromagnetic (EM) spectrum has opened up a new era in Astronomy and firmly
established the connection of binary neutron star mergers with sGRB
\citep{Abbott2017d_etal}.
This new observation has given a lot of insight in such a process and initiated 
a deeper theoretical study and extensive numerical modeling  of such systems which 
can have a very rich phenomenology due to the different binary properties,
as the massive case of GW190425 \citep{Abbott2020}, which had a total mass of 
$3.4^{+0.3}_{-0.1}\, M_{\odot}$, and there is a claim of 
an EM follow up detection \citep{Pozarenko2020}.

The important role of neutrinos in these explosions and the 
subsequent r-processing element production which will still have unique 
signaling is flourishing these last years
\citep{Sekiguchi2015,Sekiguchi2016,Palenzuela2015,Hotokezaka2015MNRAS,
Hotokezaka:2015b,Dietrich2016, Radice2016, Lehner2016, Sekiguchi2016,
  Foucart2016a, Bovard2017, Dietrich2017, Dietrich2017b, Radice2018a,
  Papenfort2018,Fernandez2018, Siegel2018}.
The amount of physics one can extract from these events is enormous. 
The possibility of a long lived remnant, its properties (see e.g.
\citep{Hanauske2016, Kastaun2016,Fujibayashi2017,Ciolfi2017,
Fujibayashi2017b})
and its consequences to electromagnetic  modeling are essential.

The main possible outcomes of a binary neutron star merger are three:
\textit{(i)} a prompt collapse to a black hole (BH),
\textit{(ii)} a neutron star that later collapses to a BH 
and \textit{(iii)} a stable neutron star.
The \textit{(ii)} scenario can be further subdivided into classes 
depending how long 
the remnant has lived. In this study we stick to the first scenario
 and will not go to more detail for the other cases 
(\cite{Baiotti2016,Nathanail2018c} for reviews).
 When the threshold mass is above a certain 
limit then this is the outcome of the merger remnant, the prompt 
collapse  to a BH 
\citep{Bauswein2013,Koeppel2019}.
The prompt collapse is very sensitive to the equation of state (EOS) 
\citep{Hotokezaka2011}.
\begin{table*}
  \centering
  \begin{tabular}{|c||c|c|c|c|c|c|}
    \hline
 model   & $B_{1,\,\rm max}$& $B_{1,\, \rm pole}$ & $B_{2,\,\rm max}$ &$A_{b,\,1}$& $A_{b,\,2}$& $E_{\rm EM}$\\
    \hline
    & $[10^{14}\, \rm G$]& $[10^{14}\,\rm G$]& $[10^{14}\,\rm G$] &$[10^{-5}]$& $[10^{-5}]$& $[10^{42}\,\rm erg$]\\
     \hline
    \hline  
	\texttt{Al$_1$} & 1.049& 0.15   & 1.049 & 2.2 & 2.2 & 7.91 \\
    \texttt{Al$_2$} & 0.104& 0.015  & 0.104 & 0.22 & 0.22 & $7.9\times 10^{-2}$ \\
    \texttt{Al$_3$} & 0.010& 0.0015 & 0.010 & 0.02 & 0.02 & $7.91\times 10^{-4}$ \\
    \texttt{Al$_4$} & 10.49&   1.5  & 10.49 & 22 & 22 & 791\\
  \texttt{Anti-al$_1$} & 1.049& 0.15  & 1.049 & 2.2 & -2.2& 48.18 \\
    \texttt{Anti-al$_1$, high-res.} & 1.049& 0.15  & 1.049 & 2.2 & -2.2& 48.18 \\
    \texttt{Anti-al$_2$} & 0.1049& 0.015  & 0.1049 & 0.22 & -0.22& 4.818 \\
    \texttt{unequal-B} & 1.049& 0.15 &    0.01 & 2.2 & 0.022& 14.06 \\

\hline
  \end{tabular}
  \caption{Initial parameters for the different models with aligned, anti-aligned 
and unequal magnetic field.$B_{1,\,\rm max}$ is the maximum value of the magnetic 
field strength inside the neutron star placed $55 \,\rm km$ at the positive part of the x-
axis and $B_{2,\,\rm max}$ the maximum of the second star placed at the same distance in 
the negative part respectively. Also, note that the value of the magnetic field strength
at the pole $B_{\rm pole}$, is an order of magnitude smaller than the maximum one 
for all models. Changing the sign of the vector potential gives rise to an 
anti-aligned dipole.}
\label{tab:initial}
\end{table*}

The case of the prompt collapse to a BH has been 
suggested not to be so exciting electromagnetically, 
firstly due to the limited amount of ejected mass, 
but also due to the small 
(sometimes negligible) accretion disk left around 
the BH \citep{Shibata06a,Baiotti08,Liu:2008xy,
Hotokezaka2011,Bauswein2013}. Furthermore, it is not 
certain yet if it can actually provide an engine for 
a  short GRB \citep{Margalit2019}. In order to model 
GRB central engines the inclusion of magnetic field
is compulsory. Magnetic field amplification, a magnetized funnel 
and a subsequent outflow are the ingredients shown to be 
important in the case that the remnant of a neutron star merger 
collapses to a BH with some delay
\citep{Rezzolla:2011,Kiuchi2014,Ruiz2016}. 

However, for the scenario 
of the prompt collapse the studies that include magnetic 
field are limited. Also, magnetic field amplification
may be suppressed in this scenario due to the quick 
production of a black hole \citep{Kiuchi2015}.
The torus formed around the BH may have a 
lifetime  as small as  
$t_{T} \sim 5 \left(\frac{M_T}{0.001M_{\odot}}\right) 
\left(\frac{\dot{M}}{0.2M_{\odot}s^{-1}}\right)^{-1} \, {\rm ms}$
 \citep{Ruiz2017a}. In this amount of time no 
 magnetized outflow is possible to be produced \citep{Ruiz2017a}.
However, in such a short timescale the magnetic energy stored 
around the neutron stars will be released in a similar way of 
a neutron star collapse \citep{Falcke2013,Most2017}. 
In an ideal MHD framework this can not be captured since 
matter is coupled to the field. The framework we use, of 
general-relativistic resistive magnetohydrodynamics (GRMHD), 
allows for electromagnetic field evolution in vacuum. 
As such the magnetic field can escape once the two stars merge 
and leave a negligible amount of mass around them.



Studying the interactions of the magnetospheres of the two 
neutron stars prior to merger can set the limit to 
electromagnetic signals before merger. 
These interactions have been studied in the case of a force-free 
realistic magnetosphere 
modeling \citep{Palenzuela2013,Palenzuela2013a, Ponce2014}. 
In our study we limit our focus to the merge and the subsequent 
evolution, since we cannot acquire reliable results prior to merger.

We begin by considering a simple
head-on collision of two magnetized neutron stars.
 Electromagnetically, 
such a scenario could have a similar signal with a prompt collapse,
the case where a neutron star merger produces a  
 remnant that quickly undergoes a gravitational collapse 
to a BH, and happens when the total mass is above a threshold 
\citep{Hotokezaka2011,Bauswein2013,Koeppel2019}. 
One of the main features affecting the pattern and EM counterpart 
of BNS mergers, is the structure of the magnetic field.
The simplistic approach we adopt here certainly misses the interesting 
aspects that would result from the dynamics of the binary 
affecting the EM field. However, we can provide a rough and conservative 
estimate of the amount of the magnetic energy, which is
stored in the two magnetospheres, and will dissipate away after 
merger due to the lack of a significant disk to hold it.
This indeed, needs to be proven from a numerical simulation 
perspective, since similar studies of BNS mergers 
undergoing prompt collapse to a BH cannot measure any EM luminosity, 
due to the adopted ideal MHD numerical scheme \citep{Ruiz2017}, 
and thus suggest that prompt collapses do not result in any EM luminosity.
The resistive MHD framework that we adopt for this study, allows 
the evolution of the EM field in vacuum and this can yield a rough 
estimate of the luminosity of such an event. 

The initial configuration that we use has a zero electric 
field in the atmosphere and is free of charges, thus suitable 
for an electrovacuum approach.
Head on collision of self gravitating stars have been studied in order to 
study the conditions of BH formation
\citep{Rezzolla2013,East2012, Kellermann:10}. 
We choose to be in the parameter space that the head on collision of two 
neutron stars would always form a BH. In our case the essential 
part is the inclusion of magnetic fields inside and outside the stars. 

We explore a magnetised neutron star head on collision as a toy model 
to understand the electromagnetic signal after a prompt collapse, where 
a negligible disk is left around the BH and as a result the magnetic 
field can escape in a millisecond timescale. 
We focus on three main models, one case where the 
dipole magnetic moments in both neutron stars are aligned, another where they 
are anti-aligned and one where the two dipoles are aligned but with one dipole 
having a maximum strength smaller by two orders of magnitude. Our goal is to 
systematically study the major differences in the evolution and the overall energetics of 
the emitted EM bursts. In the first two cases, the magnetic field is of the same 
order of magnitude in both stars.  We discuss the 
similarities of these cases with the gravitational collapse of a single neutron 
star \cite{Most2017,Nathanail2017}.
We follow the evolution after BH formation
until the EM fields decay to insignificant values,
 and further set limits on the expected flux that 
would be detected from a BNS merger that promptly collapsed to a BH with similar 
properties and distance to GW190425 \citep{Abbott2020}.

It has been proposed that NS mergers may 
give rise to fast radio bursts (FRBs) \citep{Lasky2013,
Zhang2016,Wang2016, Metzger2016,Piro2017}.
In this study we state the proposition that FRBs can be related to neutron 
star mergers only in the case that the remnant undergoes 
a prompt collapse to a BH \citep{Paschalidis2018,Nathanail2018}.
The simulations we study show that in the case of a negligible 
disk around the remnant BH, the magnetic field dissipates 
and produces giant EM pulses.
In this study we do not model the last orbits of a 
quasi-circular binary, but rather restrict only to the collision.
We believe that these results can be considered as a toy model 
that can give an illustrative picture of the EM pulses expected
from a prompt collapse to a BH.

This paper is organized as follows:
in Sec. \ref{sec:nsaid} we review the numerical setup of the simulations 
together with the initial data. In Sec. \ref{sec:mis} we present the 
numerical results and the comparison between the various models.
 In Sec. \ref{sec:pairp} the astrophysical relevance of pair production 
and magnetic reconnection is discussed, 
whereas the luminosity and the observable limits 
are left for Sec. \ref{sec:EM}. Finally, the 
discussion and the conclusions of our results are presented in Sec.
\ref{sec:con}.

\section{Numerical setup and Initial data}
\label{sec:nsaid}
%
\begin{figure}
\setlength{\tempheight}{0.22\textheight}
\centering
\rowname{\texttt{Aligned}}
\subfloat{\includegraphics[height=\tempheight]{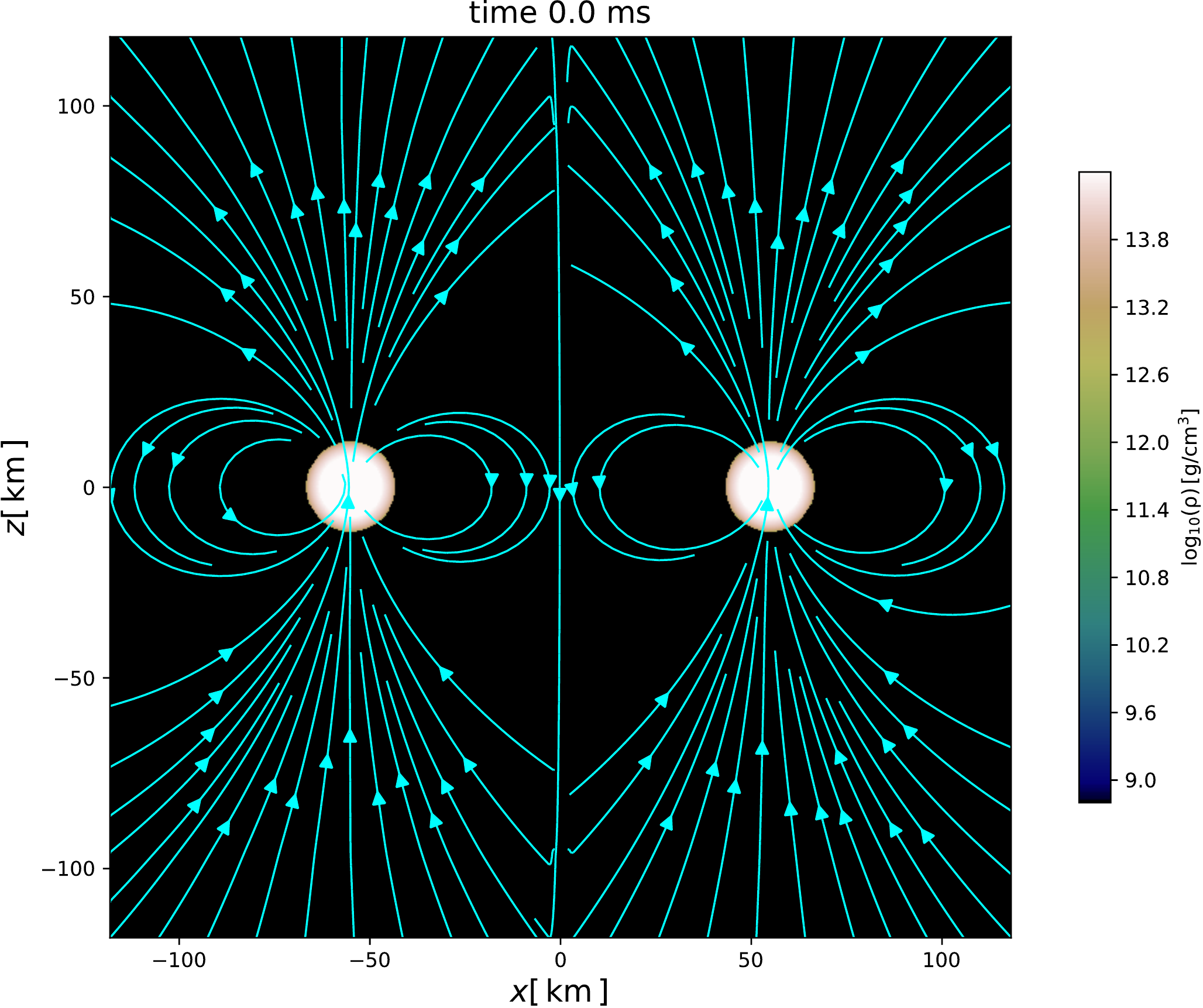}}\\
\rowname{\texttt{Anti-Aligned}}
\subfloat{\includegraphics[height=\tempheight]{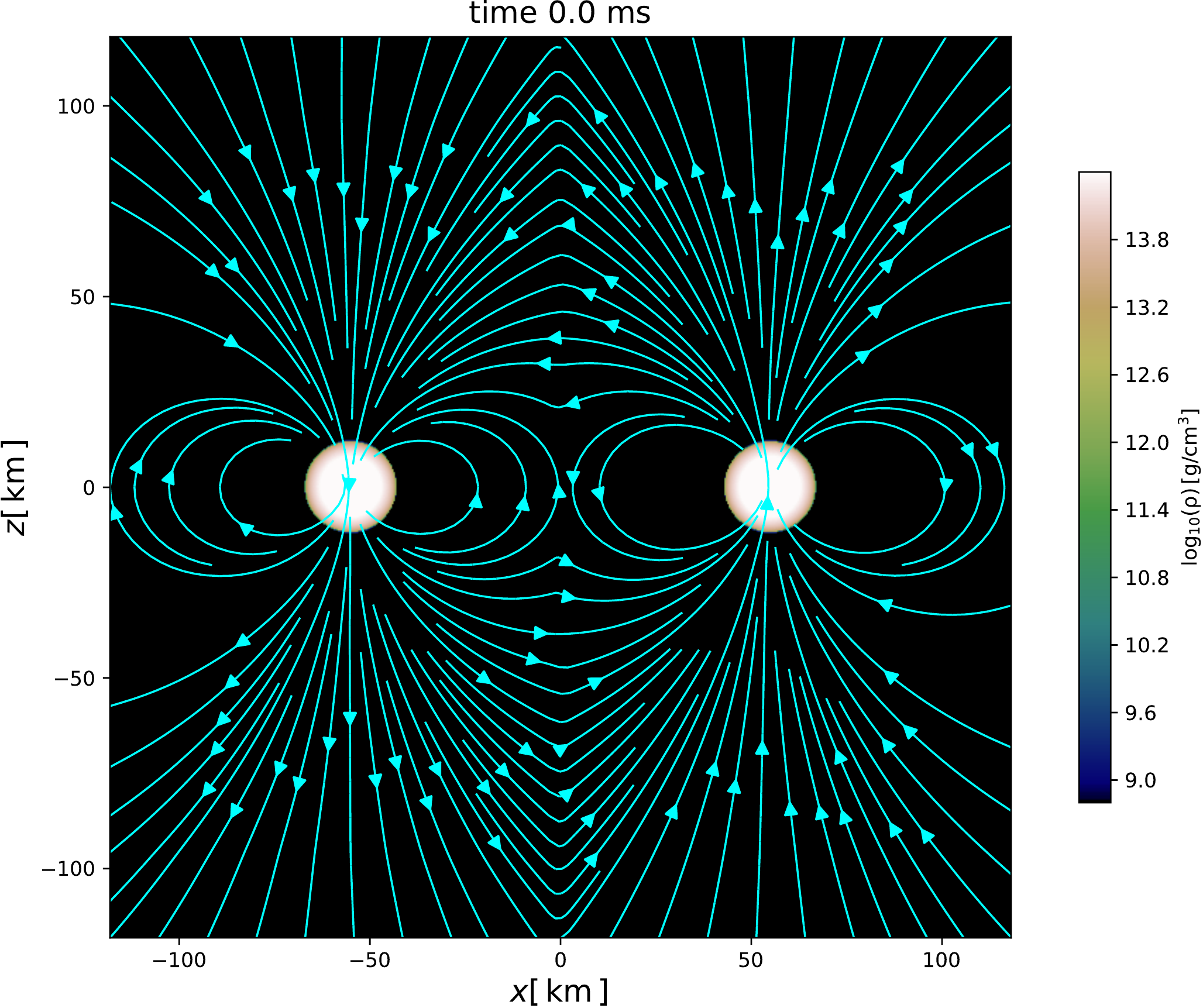}}\\
\rowname{\texttt{unequal-B}}
\subfloat{\includegraphics[height=\tempheight]{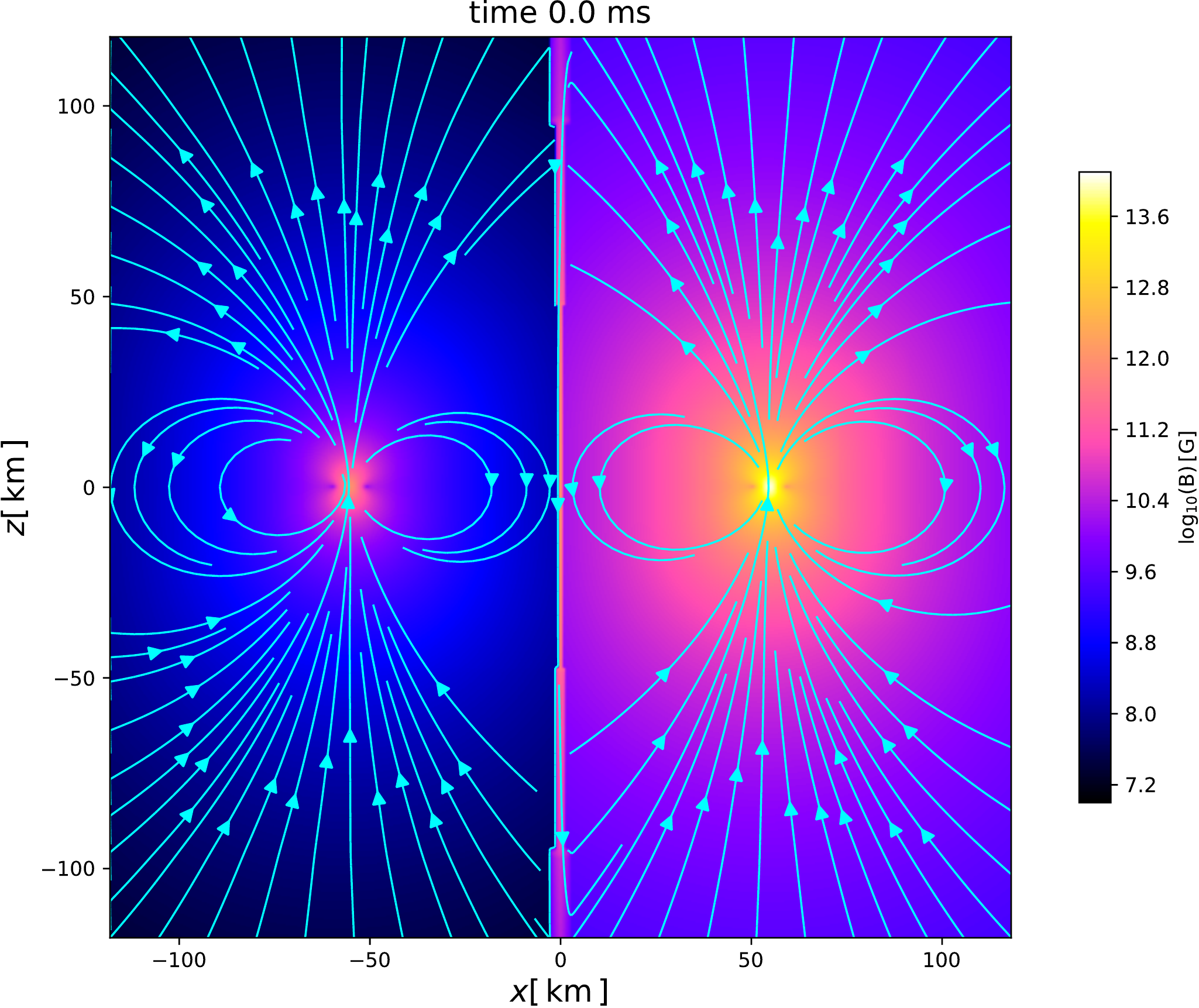}}
	\caption{Magnetic field configuration at the initial time,
	upper panel: the rest-mass density for model \texttt{Al$_1$},
	middle panel: the rest-mass density for model \texttt{Anti-al$_1$} and
	lower panel: the magnetic field  strength for model \texttt{unequal-B}, 
	is depicted since the density profile and magnetic field lines are the 
	same as in the upper panel.}
	\label{fig:mag_init}
\end{figure}

All simulations presented in this paper have been performed using the
general-relativistic resistive magnetohydrodynamics (GRMHD) code
\software{resistive WhiskyRMHD} \citep{Dionysopoulou:2012pp,Dionysopoulou2015} 
embedded in the \software{Einstein Toolkit} \citep{loeffler_2011_et}
and box-in-box mesh refinement is provided by \software{Carpet}, where 
initially tracks the two stars \citep{Schnetter-etal-03b}.
The numerical setup is similar to the one presented in \cite{Nathanail2017, Most2017}.
The GRMHD equations are solved using high-resolution shock capturing methods
like an LLF Riemann solver and the reconstruction of the primitives is done by  
enhanced piecewise parabolic method (ePPM) \citep{Colella2008, Reisswig2012b}.
The electric charge is not evolved but computed from $q= \nabla_i E^i$ at every timestep,  
similar to \citet{Dionysopoulou:2012pp} and \citet{Bucciantini2012a}. 

%
%
%
\begin{figure*}
\setlength{\tempheight}{0.20\textheight}
\centering
\textbf{density$\,\,\,\,\,\,\,\,\,\,\,\,\,\,\,\,\,\,\,\,\,\,\,\,\,\,\,\,\,\,\,\,\,\,\,\,\,\,\,\,\,\,\,\,\,\,\,\,\,\,\,\,\,\,\,\,\,\,\,\,\,\,\,\,\,\,\,\,\,\,\,\,\,\,\,\,\,\,\,\,\,\,\,\,\,\,\,\,\,\,\,\,\,\,\,\,\,|B|\,\,\,\,\,\,\,\,\,\,\,\,\,\,\,\,\,\,\,\,\,\,\,\,\,\,\,\,\,\,\,\,\,\,\,\,\,\,\,\,\,\,\,\,\,\,\,\,\,\,\,\,\,\,\,\,\,\,\,\,\,\,\,\,\,\,\,\,\,\,\,\,\,\,\,\,\,\,\,\,\,\,\,\,\,\,\,\,\,\,\,\,\,\,\,\,\,$ $|S^r|$}\par\medskip
\rowname{$t =0.3\,\rm{\rm ms}$}
\subfloat{\includegraphics[height=\tempheight]{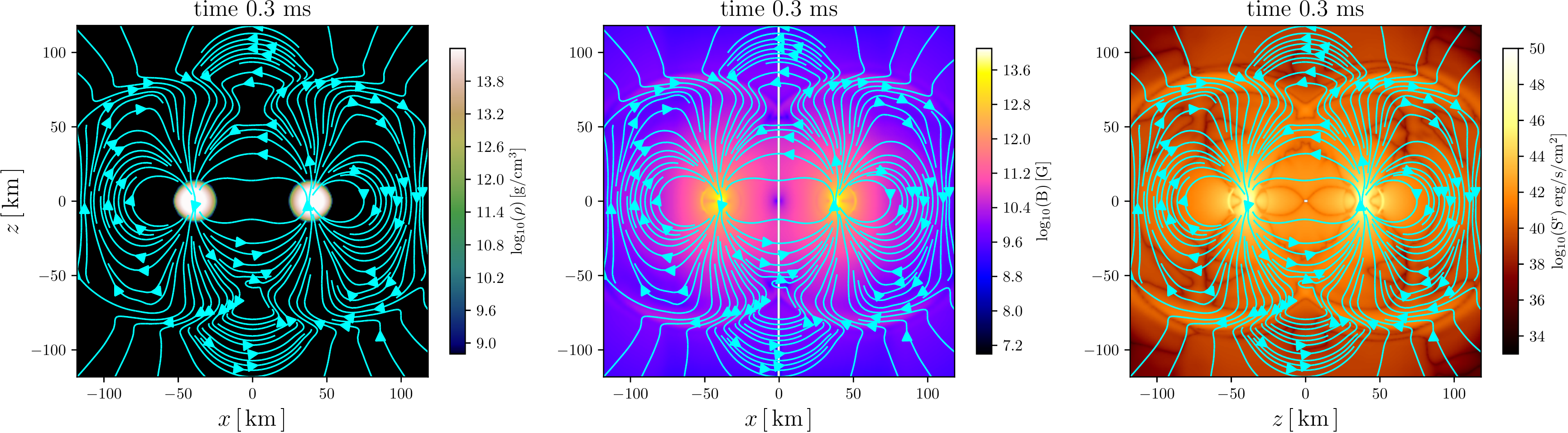}}\\
\rowname{$t =0.8\,\rm{\rm ms}$}
\subfloat{\includegraphics[height=\tempheight]{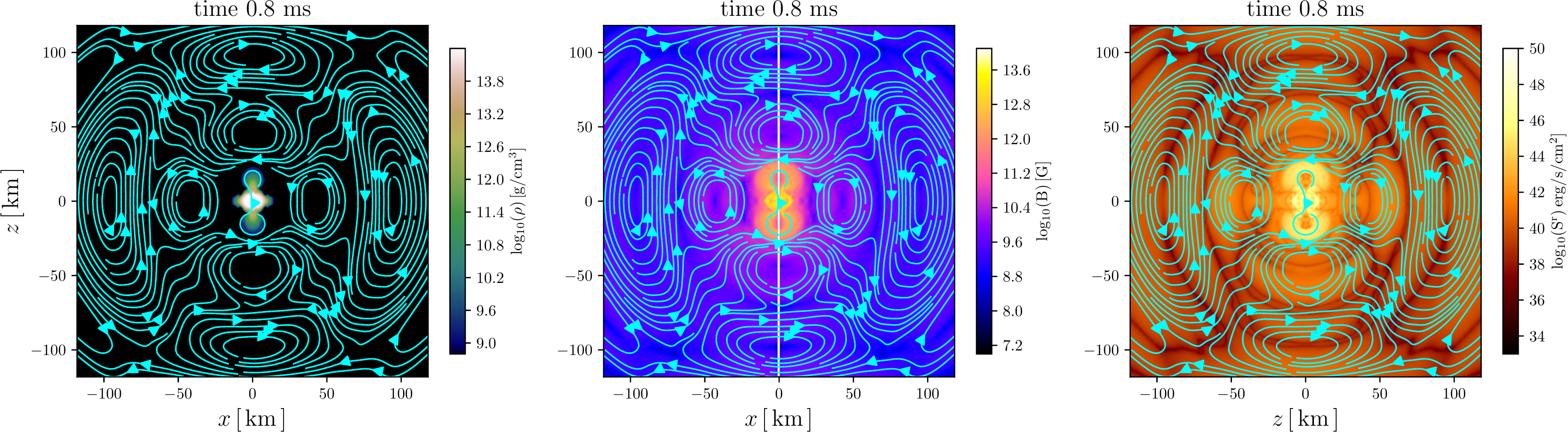}}\\
\rowname{$t =1.2\,\rm{\rm ms}$}
	\subfloat{\includegraphics[height=\tempheight]{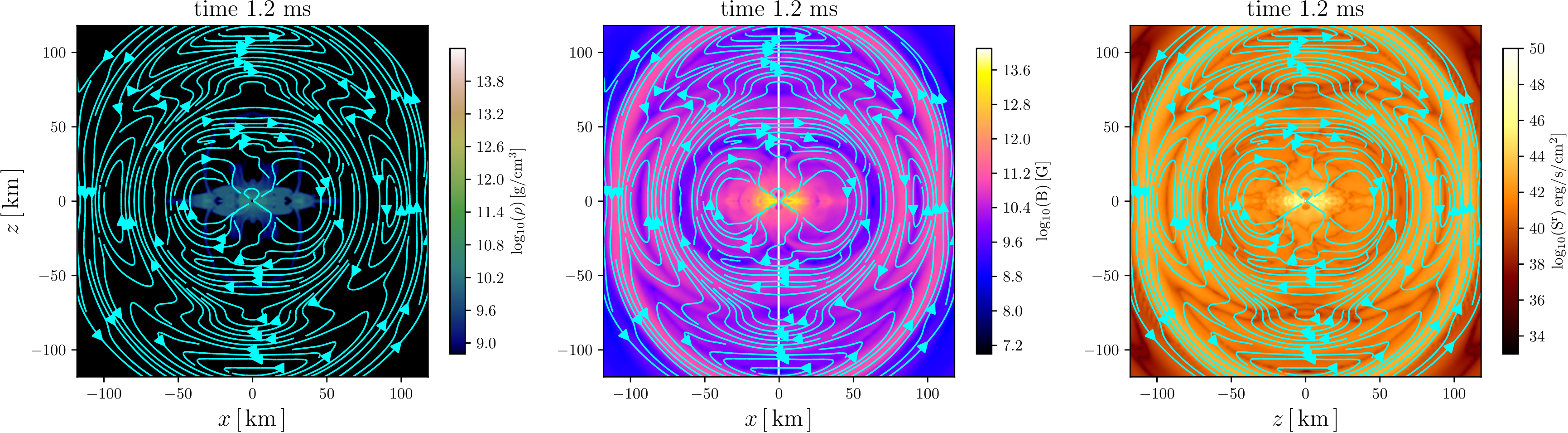}}
\caption{Anti-aligned case \texttt{Anti-al$_1$}. From left to right: evolution of the density, magnetic  
field strength and the radial component of the poynting vector $S^r$.
The different rows correspond to different times $t =0.3,\,  0.8 $ and $1.2\,\rm{\rm ms}$.}
	\label{fig:anti}
\end{figure*}
The focus of our study is the magnetic field structure and the produced 
luminosity after  the collision in which a BH is formed. 
We assume that the two stars have neutral magnetospheres,i.e, no charge density exists 
and no currents will develop in the exterior magnetospheres, and hence
model them in electrovacuum. 
The use of an ideal-MHD framework is sufficient to describe a BNS merger 
\citep{Harutyunyan2018}, but not the evolution of the magnetosphere.
In such cases we need to solve the Maxwell equations in the exterior in 
electrovacuum, which can be achieved through a resistive-MHD framework. The goal of
this approach is the inclusion of an electric current that recovers the 
ideal-MHD limit inside the star and the electrovacuum limit outside, which is filled
by a low density zero-velocity atmosphere, which then decouples from the
evolution of the EM fields.
Due to the small timescales associated with the ideal-MHD current the GRMHD equations
can become quite stiff in this limit. In order to allow for a numerically stable treatment
of these regions we employ an implicit-explicit Runge-Kutta time stepping (RKIMEX) 
\citep{pareschi_2005_ier}. For further details of our numerical setup we 
refer to \citet{Dionysopoulou:2012pp}, \citet{Dionysopoulou2015} and 
\citet{Palenzuela2013}.

To accomodate changes in the space-time due to the motion of the two neutron stars
the metric is evolved in the CCZ4 formulation \cite{Alic:2011a, Alic2013} as 
implemented by the \software{McLachlan} code \citep{loeffler_2011_et}.
One feature of this formulation is the inclusion of constraint damping terms 
\cite{Gundlach2005:constraint-damping} which can suppress violations of the 
Einstein equations and thus improve numerical stability \cite{Alic2013}.

\begin{figure*}
\setlength{\tempheight}{0.20\textheight}
\centering
\textbf{density$\,\,\,\,\,\,\,\,\,\,\,\,\,\,\,\,\,\,\,\,\,\,\,\,\,\,\,\,\,\,\,\,\,\,\,\,\,\,\,\,\,\,\,\,\,\,\,\,\,\,\,\,\,\,\,\,\,\,\,\,\,\,\,\,\,\,\,\,\,\,\,\,\,\,\,\,\,\,\,\,\,\,\,\,\,\,\,\,\,\,\,\,\,\,\,\,\,|B|\,\,\,\,\,\,\,\,\,\,\,\,\,\,\,\,\,\,\,\,\,\,\,\,\,\,\,\,\,\,\,\,\,\,\,\,\,\,\,\,\,\,\,\,\,\,\,\,\,\,\,\,\,\,\,\,\,\,\,\,\,\,\,\,\,\,\,\,\,\,\,\,\,\,\,\,\,\,\,\,\,\,\,\,\,\,\,\,\,\,\,\,\,\,\,\,\,$ $|S^r|$}\par\medskip
\rowname{$t =0.3\,\rm{\rm ms}$}
\subfloat{\includegraphics[height=\tempheight]{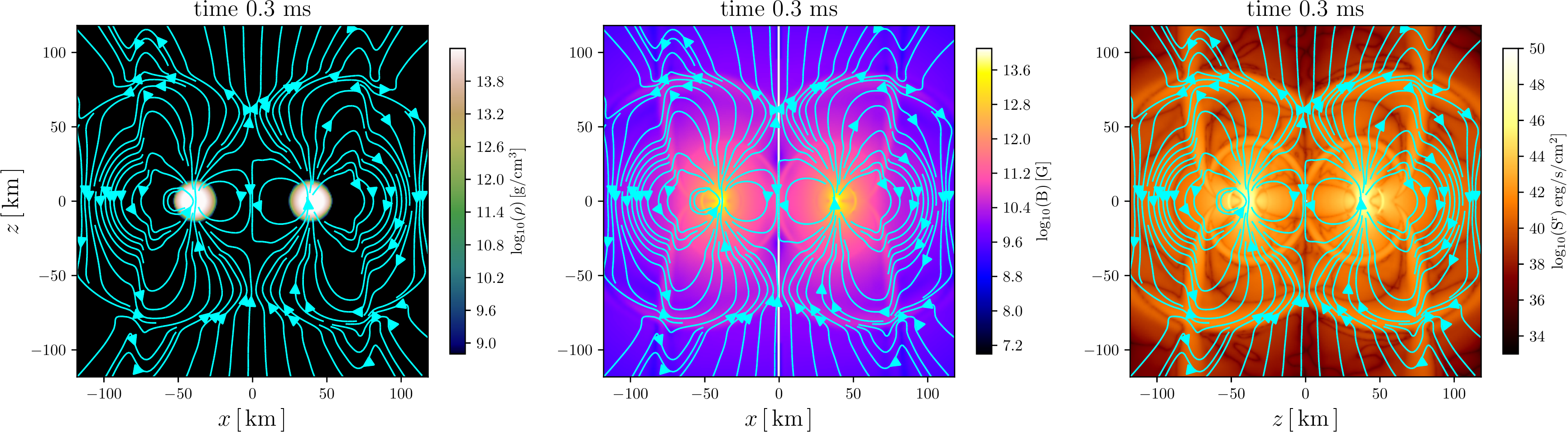}}\\
\rowname{$t =0.8\,\rm{\rm ms}$}
\subfloat{\includegraphics[height=\tempheight]{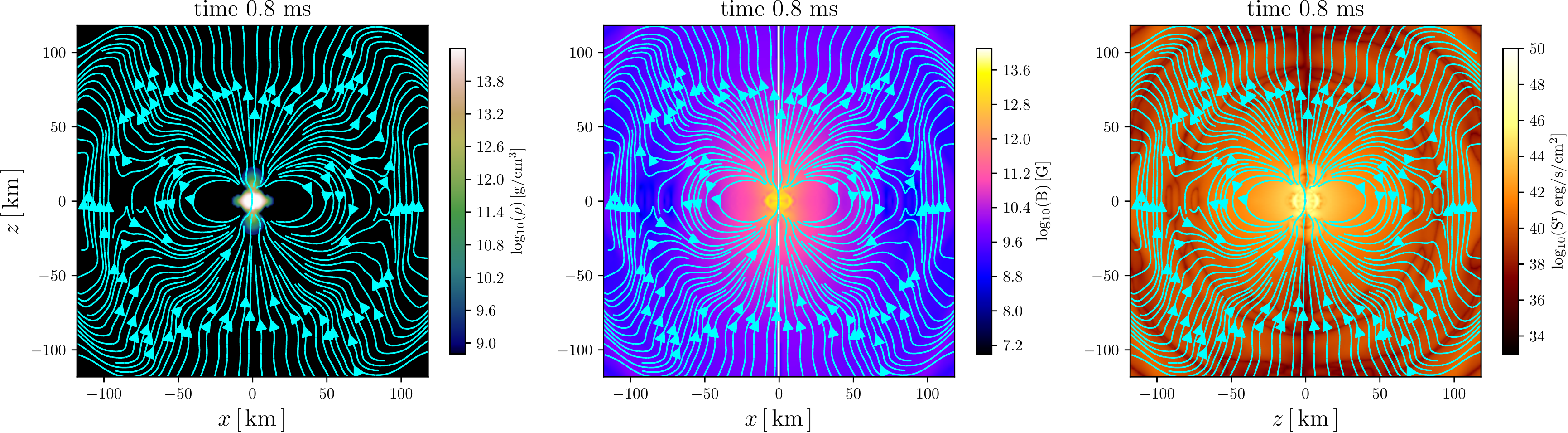}}\\
\rowname{$t =1.2\,\rm{\rm ms}$}
\subfloat{\includegraphics[height=\tempheight]{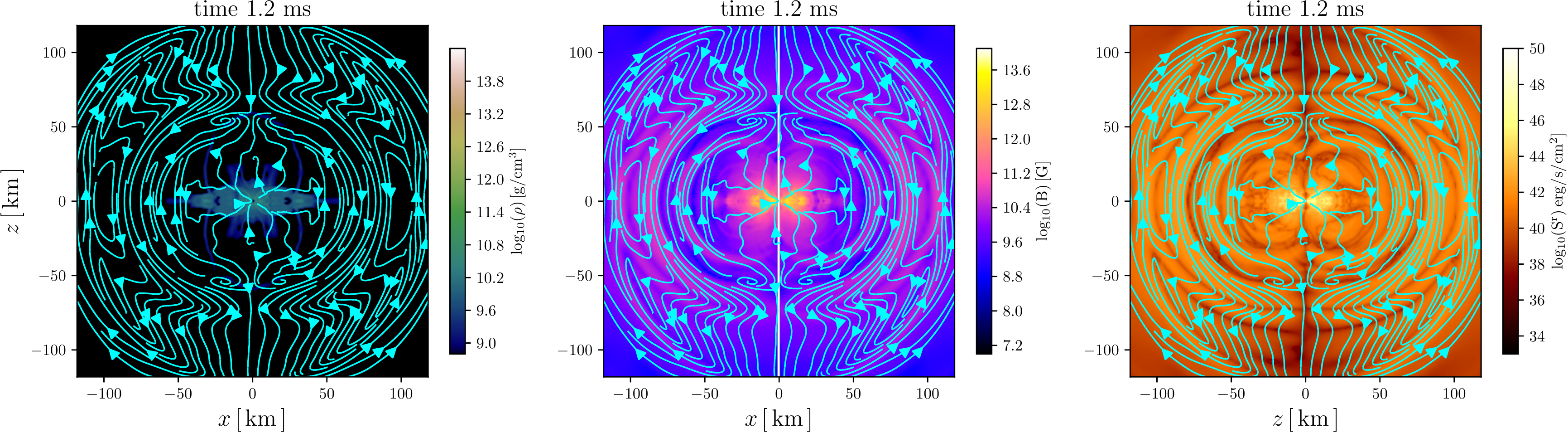}}
	\caption{Aligned case \texttt{Al$_1$}. From left to right: evolution of the density, magnetic  
field strength and the radial component of the poynting vector $S^r$.
The different rows correspond to different times $t = 0.3,\, 0.8$ and $1.2\,\rm{\rm ms}$.}
	\label{fig:al}
\end{figure*}
This is of particular importance to our choice of initial data. Previous
studies of head on collisions in pure hydrodynamics have either
resolved the constraint equations \cite{Paschalidis2011} or have 
superimposed two TOVs solved in isolation far enough apart in order 
to minimize violations \cite{Kellermann:10,Rezzolla2013}. In this study, 
we use the constraint damping terms in the
evolution scheme to remove inconsistencies in the space-time variables of our
initial data. Such an approach is in line with studies of spinning neutron stars
in binary systems \cite{Kastaun2013, Kastaun2016b} or studies of eccentric
encounters of binary neutron stars \cite{Radice2016,Papenfort2018}.

Accordingly, we consider two neutron stars separated by 110 km along the
$x$-axis. They are initially
endowed with a dipole magnetic field extending also to the exterior of the two
stars. The phi component of the vector potential, $A_{\phi}$,
 that we use to generate the dipole field is given 
below \citep{Shibata2011b}: 
\begin{equation}
A_{\phi} = \frac{A_b  ~ r_{d}/\sqrt{2}}{\sqrt[3]{(x^2+y^2+z^2+1/2 
r_{d}^2)}}\,, 
\label{Dipole}
\end{equation}
\begin{figure*}
	\begin{center}
	\includegraphics[width=\textwidth]{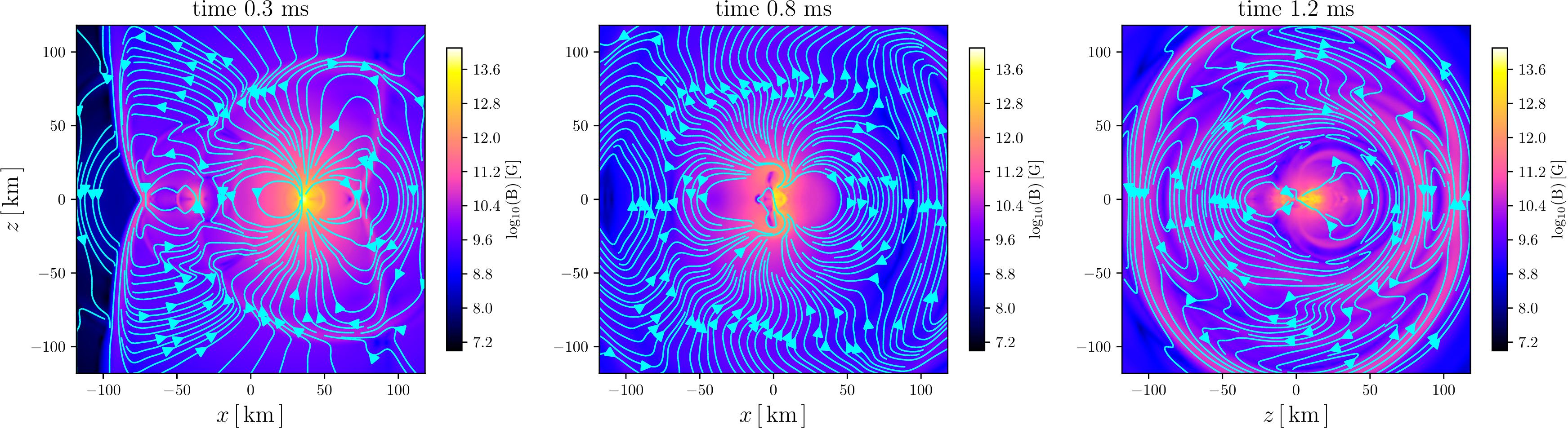}
	\includegraphics[width=\textwidth]{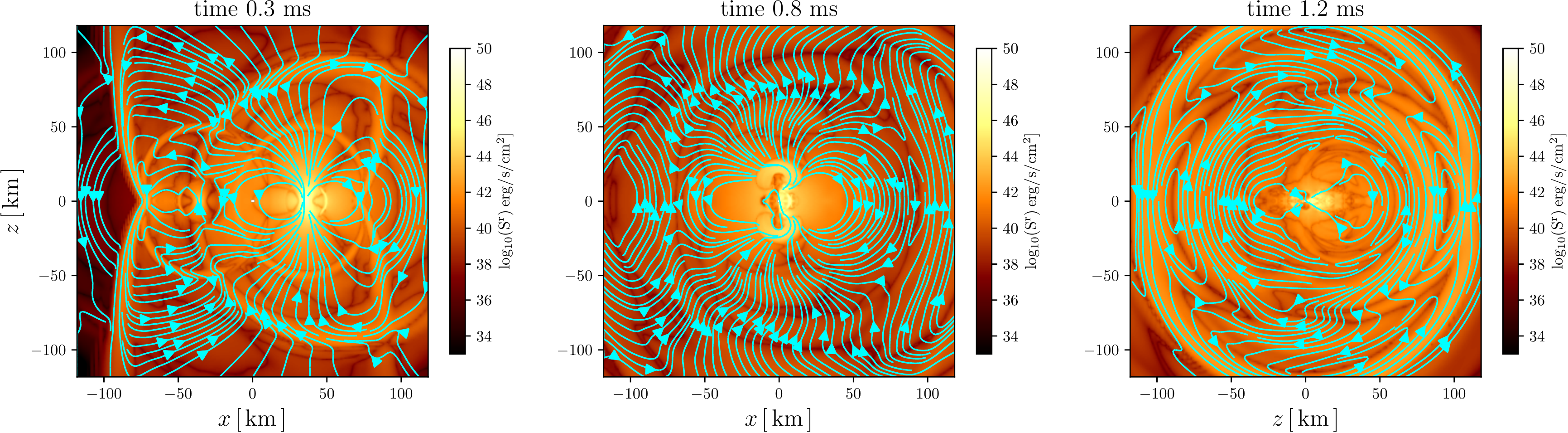}
\end{center}
	\caption{For the unequal magnetic field case, evolution 
	of the magnetic  field strength (upper panels) 
	and the	radial component of the 
	 poynting vector $S^r$ (lower panels), 
	 in the two rows respectively. The different 
	 columns correspond to different times at $t = 0.3, 0.8$ and $1.2{\rm ms}$.}
	\label{fig:un-B}
\end{figure*}
where $r_{d}$ is the radius of the current loop that generates the dipole and 
$A_b$ is a scaling factor that determines the strength of the field.
We consider two different cases for the magnetic field initial geometry, 
one where both dipoles are aligned with respect to each other and the symmetry 
plane ($y-z$) and another where one dipole is anti-aligned with the other.
Notice that by changing the sign of the vector potential we get the  anti-aligned 
magnetic field. Every magnetosphere occupies the half domain and there 
is a mismatch exactly at the $y-z$ plane, which passes through the origin.
The reason why we choose this configuration, is to allow for an initial readjustment
of the magnetospheres as the evolution begins, instead of an initial superposition
of the two dipoles. The choice we make results in an initial peak in 
radiation, analogous to the so-called "junk radiation", which quickly reduces 
to two orders of magnitude, before the actual burst from 
merger is detected, this is discussed with the presentation of fig. 
\ref{fig:poynt}. The neutron stars have an initial 
separation of $110\, {\rm km}$, in order to allow for the impact of the 
initial "junk radiation" to decrease significantly (two orders of magnitude 
in luminosity), before the main EM emission is produced by the merger itself.
Furthermore notice that due to the lack of a quasi-circular orbit, the merger 
occurs in less than a ${\rm ms}$. The initial configuration can be seen in 
fig. \ref{fig:mag_init} where we plot the 
magnetic field lines at the initial setup for the three 
representative models.

%

%

Our numerical domain consists of six refinement boxes, where the resolution 
doubles when going to the next higher refinement level. The outer boundary is placed at 
$\sim 378$ km. As the two stars move towards each other, the finest resolution 
is $\Delta x \sim 367$ m. At the time of merger, where the collapse is 
triggered, an extra refinement level is added making the highest 
resolution $\Delta x \sim 183$ m, note that this is considerably higher than 
the one used in \cite{Palenzuela2013a}. 
In order to check the robustness of our results with respect to resolution, 
we run an extra simulation, model \texttt{Anti-Al$_1$, high-res.} with the finest 
resolution at $\Delta x \sim 294$ m, which at the time merger becomes
$\Delta x \sim 147$ m.

The neutron stars are modeled as non-rotating. We have considered 
a simple polytrope with $\Gamma 
= 2$ and $K = 100$ which for a gravitational mass of $M=1.4 M_{\odot}$
yields a radius $R_{NS}=11.94$ km and central density of $\rho_c = 7.92\times 10^{14} 
~ {\rm g/cm^3}$. The choice of equation of state in this study is not so 
relevant, since our focus is on the EM pulses produced by the magnetic field 
which decouples from the fluid and dissipates away.
Both stars have an initial velocity $v^x  \simeq 0.15 c$, moving 
on the $x$-axis towards each other. 


%
\section{Numerical results}
\label{sec:mis}
 %
\begin{figure*}
	\begin{center}
		\includegraphics[width=\textwidth]{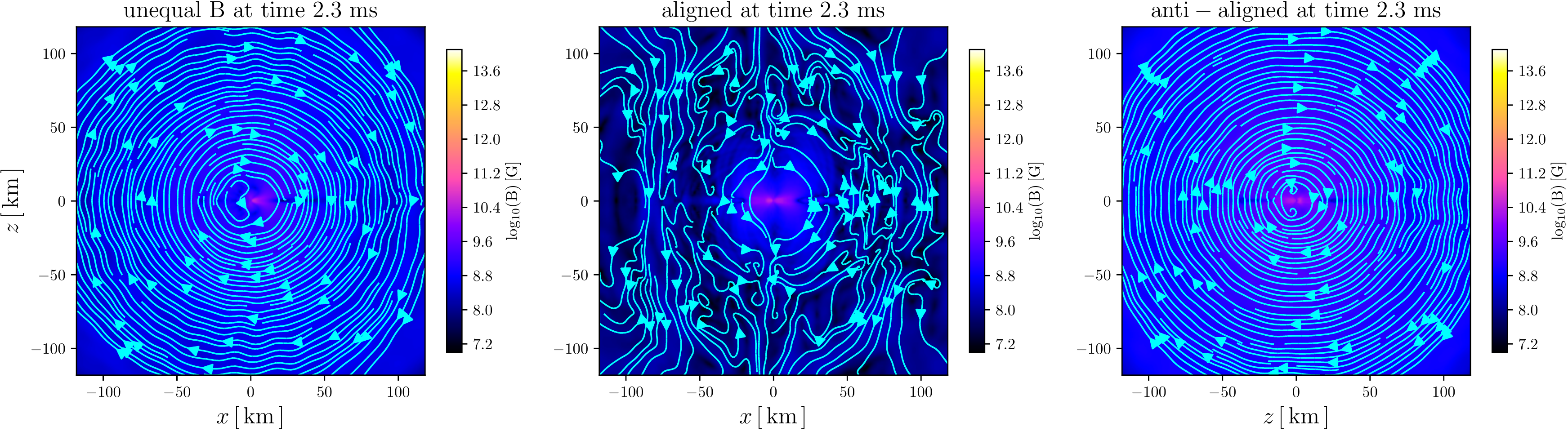}		
\end{center}
	\caption{Magnetic field strength for for three different models at 
	late time $t=2.2 sec$. From left to right is the unequal B 
	(\texttt{$unequal-B$}), aligned (\texttt{$Al_1$}) and
	 anti-aligned (\texttt{$Anti-al1$}) case respectively. 
	 The maximum magnetic field strength in 
	 the vicinity of the BH is less than $10^9 G$.}
	\label{fig:late}
\end{figure*}
In this section we will present and describe the results of our numerical 
simulations
for the head on collision of two magnetised neutron stars for six 
different cases, which are briefly described in Table \ref{tab:initial}.
The differences in the models used in this study are purely in the 
magnetic field configuration. All the hydro-dynamical properties of the 
stars are the same in all runs. The different magnetic field configurations 
are as follows: four cases for two aligned dipoles with the same 
maximum magnetic field strength, which ranges $B_{\rm max}\simeq 10^{12},
\, 10^{13}, \, 10^{14}$ and $10^{15} \, \rm G$ respectively and 
one case where the two magnetic fields differ by two orders of magnitude, namely the 
one star has $B_{\rm max}\simeq 10^{14}\, \rm G$ and the other 
$B_{\rm max}\simeq 10^{12}\, \rm G$. And lastly, two cases
 with anti-aligned magnetic dipoles 
 with  magnitude of $B_{\rm max}\simeq 10^{13}$ and $10^{14}\, \rm G$
 respectively. 
The reason to choose four different magnetic field strengths for 
a single model is to acquire and test the expected relation between 
magnetic field strength and luminosity, further note that since the 
density scaling in numerical relativity simulations is tied by 
the construction of the neutron star, every such simulation has a different 
value for the plasma beta parameter inside the star, $\beta=p/p_m$ 
(fluid pressure $p$ over magnetic pressure $p_m$).
This scaling is discussed and recovered in Sec. \ref{sec:EM}, and thus 
the results from the other models can be extrapolated according 
to this scaling, for different magnetic field strengths.

%

The matter and space-time dynamics in 
a head-on collision of two NS have been studied in detail \citep{Kellermann:10, East2012, Rezzolla2013}.
We only sketch briefly the evolution of the matter which is essentially 
identical in all cases. The two NS move towards each other 
with an initial velocity of $v^x  \simeq 0.15 c$. 
The two stars touch at time $t_{\rm merge} \simeq 0.58 \, {\rm ms}$, 
an apparent horizon is found for the first time at $t_{\rm BH} \simeq 0.81 {\rm ms}$
(fig. \ref{fig:anti} and \ref{fig:al})
and the bulk of the matter has already crossed the horizon 
by $t \simeq 1.7\, {\rm ms}$.
 
The first case that will be described is the one where initially the 
two NS are endowed with a dipolar field anti-aligned with respect to 
each other, model \texttt{Anti-al$_1$}.
When the stars start to move, the two magnetospheres begin to interact. The 
opposite field components cancel out and the field lines reconnect,
as can be seen in the first row of fig. \ref{fig:anti}. As both 
stars come closer this structure evolves to a quadropolar-like field
(second row fig. \ref{fig:anti}). 
On both sides right and left the dipole field of each star is still 
well structured and with different polarity with respect to each other.

At the time of merger ($t\simeq 0.58 \,\rm{\rm ms}$) these 
magnetic-loop structures are still anchored on the stellar matter. 
Quickly after the two stars merge, the magnetic-loop structure is 
disengaged from the matter and radiated away. As can be seen in 
the second row of fig. \ref{fig:anti}, 
this is produced during merger and subsequently radiated away following the 
third row of fig. \ref{fig:anti}. This 
EM pattern is evident when looking at the radial component of 
the poynting vector ($S^r$, third column in fig. \ref{fig:anti} 
and \ref{fig:al}). 

After that, the explosion has a spherical shape with an excess 
of energy radiated towards 
the equator. The formed BH, following the collision of the 
two stars, continues to ring down as it settles down. This continuously 
distorts the magnetic field at the vicinity of the BH producing 
more pulses with less intensity. At time $\sim 1.6{\rm ms}$ the magnetic field 
strength around the BH is less than $10^{10} G $ and 
continuous to decay. The loop structure of the magnetic field is still being 
distorted by the settling of the BH. 


The second case that we study, is the one with both dipoles aligned and 
parallel to the $z$-axis (model \texttt{$Al_1$}, from Table \ref{tab:initial}).  
As before, each dipole is extended and filling 
the half  space depending on where the neutron star is placed, right or 
left of the $y-z$ plane. When the simulation starts both dipoles meet exactly 
at this plane. A small shock is induced to both magnetospheres as the two 
stars start to move (first row of fig. \ref{fig:al}), which is analogous 
to the "junk radiation" reported in the literature. Note that this initial 
transient reduces significantly, before the burst from merger is detected.
 As the two stars 
are ready to merge, the magnetic structure resembles that of a big 
dipole (second row of fig. \ref{fig:al}). 
The continuation of this model closely resembles the collapse of a massive 
neutron star \citep{Most2017}. After merge, the apparent horizon is formed 
and the field lines are violently snapped as the highly conducting matter 
hides behind the horizon. During this, quadrupolar EM radiation is 
generated and propagates outwards. This quadrupolar pattern is the  main 
difference from the anti-aligned case, where the EM radiation is mostly 
spherical at this stage. In the last phase, when the bulk of the matter 
is already behind the horizon,the radiation pattern follows the 
ringing down of the BH,
we will come again to this point when analyzing the EM pulses.

\begin{figure*}
	\begin{center}	
	\includegraphics[width=0.97\textwidth]{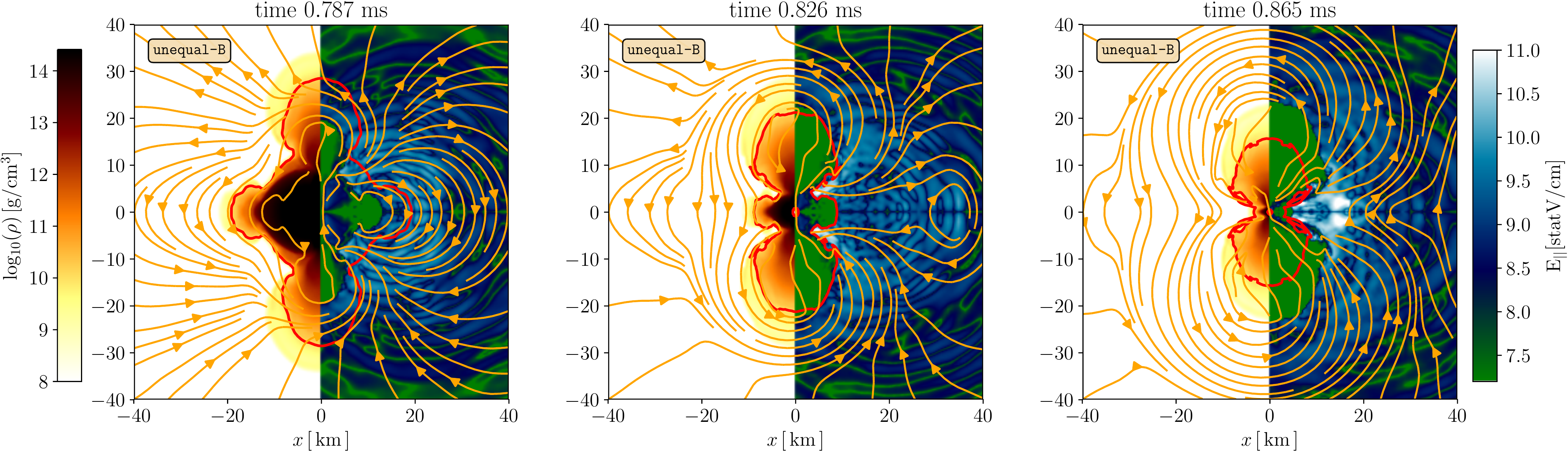}
	\includegraphics[width=0.97\textwidth]{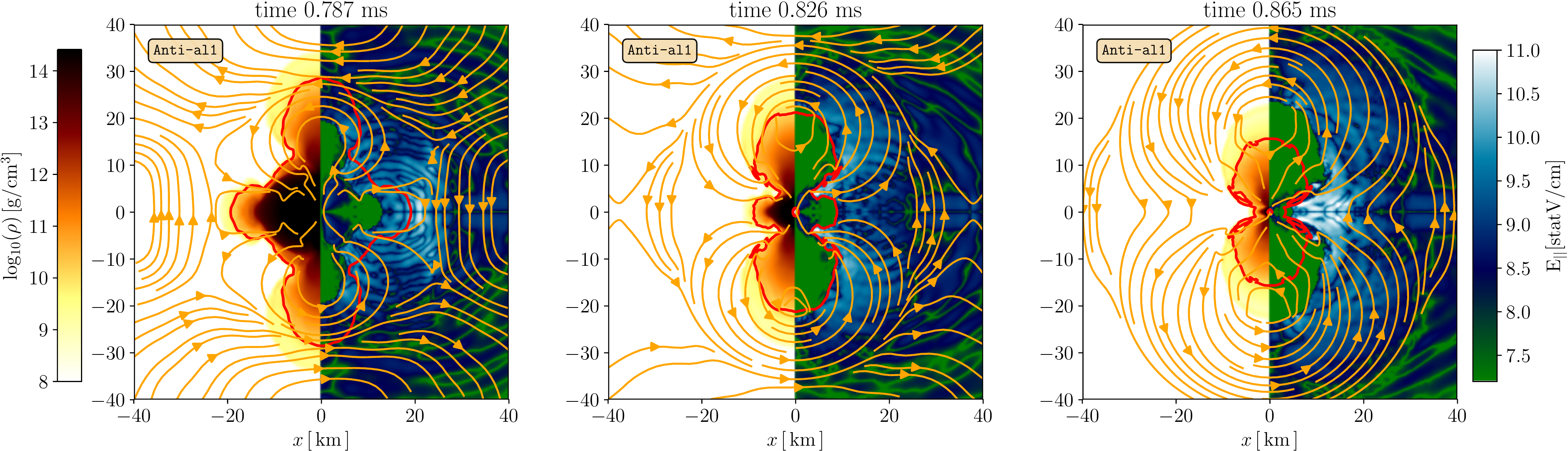}
	\end{center}
	\caption{Upper	 panels: \texttt{unequal-B}, lower panels:
	\texttt{\texttt{Anti-al$_1$}}. Left plots show rest-mass density, 
	whereas the right plots show the electric field 
	parallel to the magnetic field, both in the $(x,z)$ plane shown at three
  different times. Also reported are the 
  stellar surface (solid red line) and the 
  magnetic-field lines (orange lines).}
	\label{fig:Epar}
\end{figure*}
The next case we discuss is the model \texttt{$unequal-B$}, where the two 
NSs have aligned magnetic dipoles with a two orders of magnitude 
difference in strength, this model is the most likely 
relevant for astrophysically realistic conditions. 
 In the initial phase of this run, the magnetosphere
of the high-field NS (which is the one on the positive part of the 
x-axis \ref{fig:un-B}) quickly engulfs the magnetosphere of the 
low-field NS. This acts as a slight shock to the magnetosphere 
of the high-field NS. At the time of merger and after that, the EM radiation
follows a similar pattern with that of model \texttt{$Al_1$}, the main
difference is that the right (undistorted) side is more efficient 
in radiation as can be seen in the third column of fig. \ref{fig:un-B}.
The main features of the magnetic field follow the evolution of the high field 
magnetosphere, whereas the other NS acts as a simple companion star, 
introducing a large perturbation to the overall magnetospheric structure.
Due to the resistivity of the secondary neutron star (the one with the 
lower magnetic field), the magnetic field lines are anchored on the surface 
of the star (left panel of fig. \ref{fig:un-B}). However, 
after merger the evolution of the EM field is dictated 
by the matter dynamics and the produced magnetic loops are generated 
similarly to model \texttt{Anti-al$_1$} (third column of fig. 
\ref{fig:anti} and \ref{fig:un-B}). 

It is interesting to see what is left around the BH at late 
times ($t\simeq 2.3{\rm ms}$). The case with the initial aligned dipoles 
(\texttt{$Al_1$}) closely compares with the collapse of 
one massive magnetised  NS (middle panel of fig. \ref{fig:late}, compare 
with \cite{Most2017}). For the other two cases, the anti-aligned dipoles and 
the unequal magnetic field, the late time evolution is similar, 
making a bubble-like structure around the BH. At such late times the 
strength of the magnetic field, in the vicinity of the BH, 
is less than $\sim 10^{11} G$. It is important to check and discuss 
the physical mechanism of pair creation and magnetic recconection 
in this numerical setup, as these   are expected to happen similarly 
with pulsar magnetospheres.

\section{Pair production, magnetic reconnection and high energy radiation}
\label{sec:pairp}

Before computing the EM output from the the models described in this 
study, we focus in the efficiency of pair production during 
such cataclysmic events. Moreover, we follow regions where this 
may occur and regions of magnetic field alternating polarity 
that reconnection is expected in a realistic environment.

One of the dominant mechanisms for pair creation in a pulsar magnetosphere
is the interaction of high energy (usually curvature photons) with strong 
magnetic field. The important ingredient for this mechanism to start 
to operate is a strong electric field component parallel to the magnetic 
field. In this section we discuss the occurrence  of the physical 
conditions that would trigger pair creation during an event 
similar to the ones described in these simulations. 
The initial conditions that we employ have an empty of charges 
environment around the two neutron stars. During the simulation high electric 
field is generated which we follow in order to check 
where the condition of pair creation is met.

Due to the developed voltage drop $\Delta V$, a charge attain a Lorentz factor
$\gamma$, which is given by
\begin{align}
  \gamma= e\, \Delta V / m_e \, c^2 \, ,
  \label{eqn:gamma}
\end{align}
where $e$ and $m_e$ is the electric charge and the mass of the
electron, respectively.

The condition for pair creation is given as follows:
\citep{Sturrock1971, Ruderman1975} 
\begin{align}
  \gamma^3 \left(\frac{\hbar c}{2 m_e r_c c^2} \sin \theta\right)
  \left(\frac{B_{\rm loc}}{B_{\rm cr}} \right) 
  \simeq \frac{1}{15}   \, ,
  \label{eqn:pp}
\end{align}
where $B_{\rm loc}$ is the local magnetic field, $B_{\rm cr}:= 4.4
\times 10^{13} \, {\rm G}$ is the so-called critical magnetic-field
strength and $\sin(\theta)$ is the ``pitch'' angle between the 
photon and the magnetic-field line. $r_c$ is the radius of curvature 
of the magnetic-field line that
the charge will travel on. Using the above mentioned equations, 
we report  	the criterion for triggering pair creation
\begin{align}
 E < E_{\rm pp} \simeq \, 1.5 \times 10^{11} & \left( 
 \frac{r_c}{20\, {\rm km}} \right)^{2/3}    \nonumber \\
 & \left( \frac{B_{\rm loc}}{10^{10}\, {\rm G}} \right)^{-1/3}
  \quad {\rm statV/cm }\, .
 \label{eqn:electr}
\end{align}
%

%
\begin{figure}
	\begin{center}	
	\includegraphics[width=0.97\columnwidth]{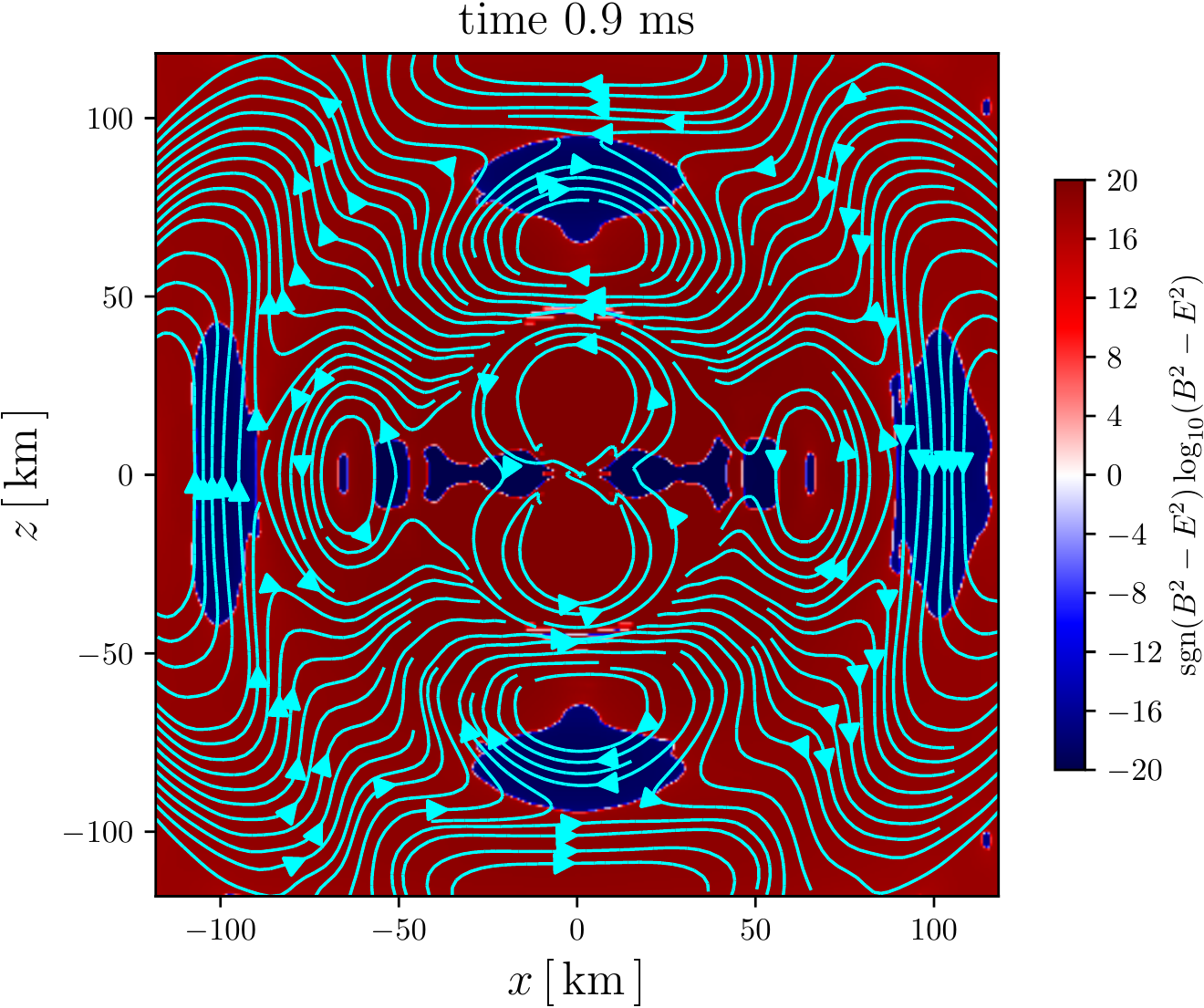}
	\end{center}
	\caption{The logarithm of $B^2-E^2$ multiplied by its sign, 
	to account for negative values,
	 in the $(x,z)$ plane shown at time $t=0.9\, {\rm ms}$
	 for model \texttt{Anti-al$_1$}. Red regions indicate magnetically 
	 dominated regions, whereas blue regions electrically dominated regions 
	 which are allowed to develop with our numerical scheme. 
	 It is exactly at the interface
	 of these two regions that magnetic reconnection is 
	 expected and  could provide energy 
	 dissipation and high energy radiation.}
	\label{fig:B2E2}
\end{figure}
In order to evaluate the occurrence of pair creation in the 
cataclysmic environment of colliding neutron stars, we report 
the strength of the parallel electric field 
$\boldsymbol{E}_{||} := \boldsymbol{E} \cdot
\boldsymbol{B}/|\boldsymbol{B}|$. The models under investigation 
are \texttt{$unequal-B$}, where the two stars have an unequal 
initial magnetic field and \texttt{Anti-al$_1$}, initial magnetic dipoles 
are anti-aligned.

In fig. \ref{fig:Epar} we show for these two models, 
 the evolution of the rest-mass density 
(left panels) and the parallel electric field
(right panels) at three representative times, at which 
$\boldsymbol{E}_{||}$ reaches a rather high value. Also shown in
fig. \ref{fig:Epar} are the stellar surface (solid red line ) 
and the magnetic-field lines (orange lines).

Note that initially, for model \texttt{$unequal-B$} (upper panels)
  the neutron star with the higher magnetic field
is placed on the positive x-axis, which means it is on the right side 
of the figures. This is the reason of the asymmetry of the magnetic field
lines as can be seen in all panels.
This is also imprinted in the development of the 
parallel electric field where on the right side it reaches the 
highest values, whereas on the left it retains low values at all times.
As a result pair creation  
would be expected in a neutron star head on collision, if one of the two 
stars has the limiting initial magnetic field of 
$B_{\rm  pol} = 10^{13}\,{\rm G}$, lower values would most 
probably suppress  pair creation. 

For model \texttt{Anti-al$_1$} (lower panels of fig. \ref{fig:Epar}), 
a similar situation with model \texttt{$unequal-B$} is depicted, 
high values of electric field are generated at regions 
where the magnetic field is changing polarity (visible in all three panels). 
Matter dynamics, during merger and close to the collapse to a BH, 
twist the magnetic field lines generating an enormous electric field in the 
region where magnetic reconnection is expected. In the middle and rightmost 
plots of the lower panels of fig. \ref{fig:un-B}, at the left part 
of the plots near the equatorial plane, it is seen that the magnetic field 
is changing polarity and exactly at this place a huge parallel 
electric field is developed. From our simulations we can extract these estimates 
for the strength of the developed parallel electric field, since we allow 
regions with $B^2-E^2<0$ to develop. This is also illustrated 
in fig. \ref{fig:B2E2} were we plot the quantity $B^2-E^2$ in a logarithmic 
scale, but also incorporating the sign of it, to better understand these 
regions were electric fields develop and magnetic reconnection is 
expected in a realistic environment. Regions of magnetic field alternating 
polarity can be observed at the interfaces of magnetic loops with different 
orientation, in fig. \ref{fig:B2E2} these are located on a circle 
of radius $\sim 100\, {\rm km}$. Moreover, similar regions of 
alternating magnetic field polarity are observed on the 
equatorial plane from $\sim 50\, {\rm km}$ till the star surface, 
which are seen  in fig. \ref{fig:B2E2}. 

In our simulations the 
EM field energy is carried away by these huge magnetic loops described 
above, these magnetic loops are generated from the merger of the two 
magnetospheres and are further influenced by the dynamics 
of the matter, which after merger have this bubble  shape 
shown in the left-part of all plots in fig. \ref{fig:Epar}.
The magnetic bubbles that are produced, move outwards in all directions, 
from the equatorial plane to the north and south, and let us estimate 
the magnetic energy dissipated after merger. Under the conventions of the 
numerical scheme used in our simulations no dissipation can be modeled 
locally, but only a global picture can be drown, of the 
magnetic energy  that escapes after merger in the form 
of Poynting flux. As such, the estimations derived in the next section 
can be regarded as rough limits of the emission produced from 
a BNS merger that undergoes a prompt collapse.

%
\section{Electromagnetic output}
\label{sec:EM}
%
\begin{figure}
	\begin{center}	
	\includegraphics[width=0.96\columnwidth]{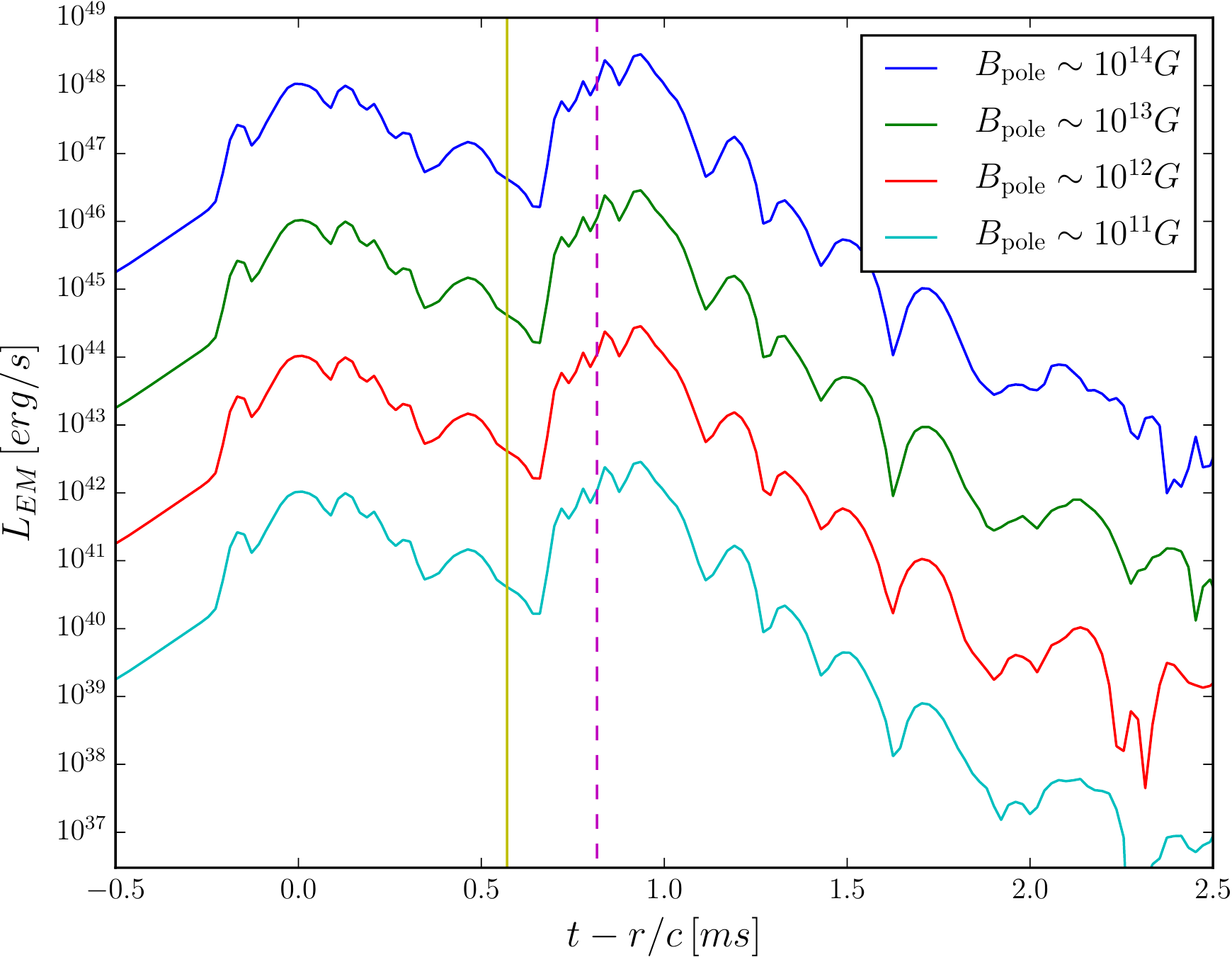}
	\includegraphics[width=0.96\columnwidth]{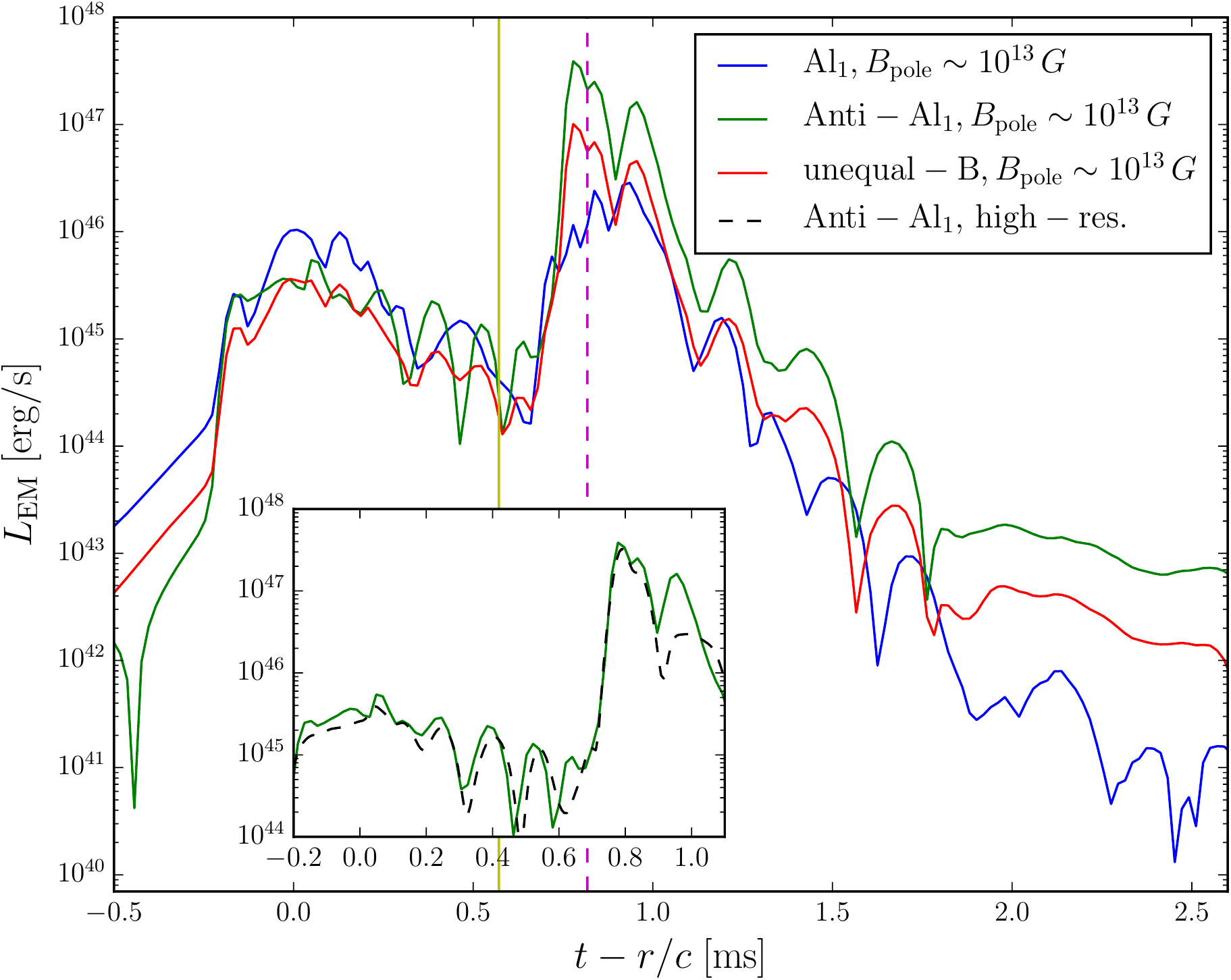}
	\end{center}
	\caption{EM luminosity is extracted at $220.5\,{\rm km}$, and
	expressed in retarded time,
	for models, upper panel: 
	\texttt{$Al_1$}, \texttt{$Al_2$},\texttt{$Al_3$} and \texttt{$Al_4$}, 
	lower panel:
	 \texttt{$Anti-al1$}, \texttt{$unequal-B$} and
	\texttt{$Al_1$}. The solid yellow line represents the time of merger and the 
	dashed magenta line the time when the apparent horizon is found for the first 
	time. The maximum magnetic field of each model is stated inside the plots. 
	In the embedded plot the \texttt{$Anti-al1$} high resolution run is compared 
	with the base resolution. }
	\label{fig:poynt}
\end{figure}
In this section we will discuss the EM emission of all the head on
collisions performed in this study.  While we are interested in giving an
estimate of the radiated energy we also track the pattern of the EM pulses
produced during such collisions. 
We compute the EM luminosity as: 
\begin{align} 
  L_\mathrm{EM} = \oint_{r=\mathrm{const}} T^{t}_{\mathrm{EM}\ r}\ \mathrm{d} \Sigma \label{figpoynting:}
\end{align}
on a surface at $r\simeq 220.5$ km from
the merger site of the two neutron stars. Before the two stars merge, the
EM signal is dominated by an early transient as the two stars start to move
towards each other, the "junk radiation" already discussed.  
This transient is the result of our ad-hoc placement
of the initial dipole magnetic fields that initially have to adapt to our
prescription for the resistive surface layer of the two stars
\citep{Dionysopoulou:2012pp}.  While EM outflows are expected from the
quasi-circular orbital motion of realistic binaries \citep{Hansen2001,
Lai2012,Piro2012}, the
transient found here is a pure numerical artifact that, however, due to
similar energetics is of similar magnitude to that expected for realistic
BNS.  
 
\begin{figure}
	\begin{center}	
	\includegraphics[width=1.0\columnwidth]{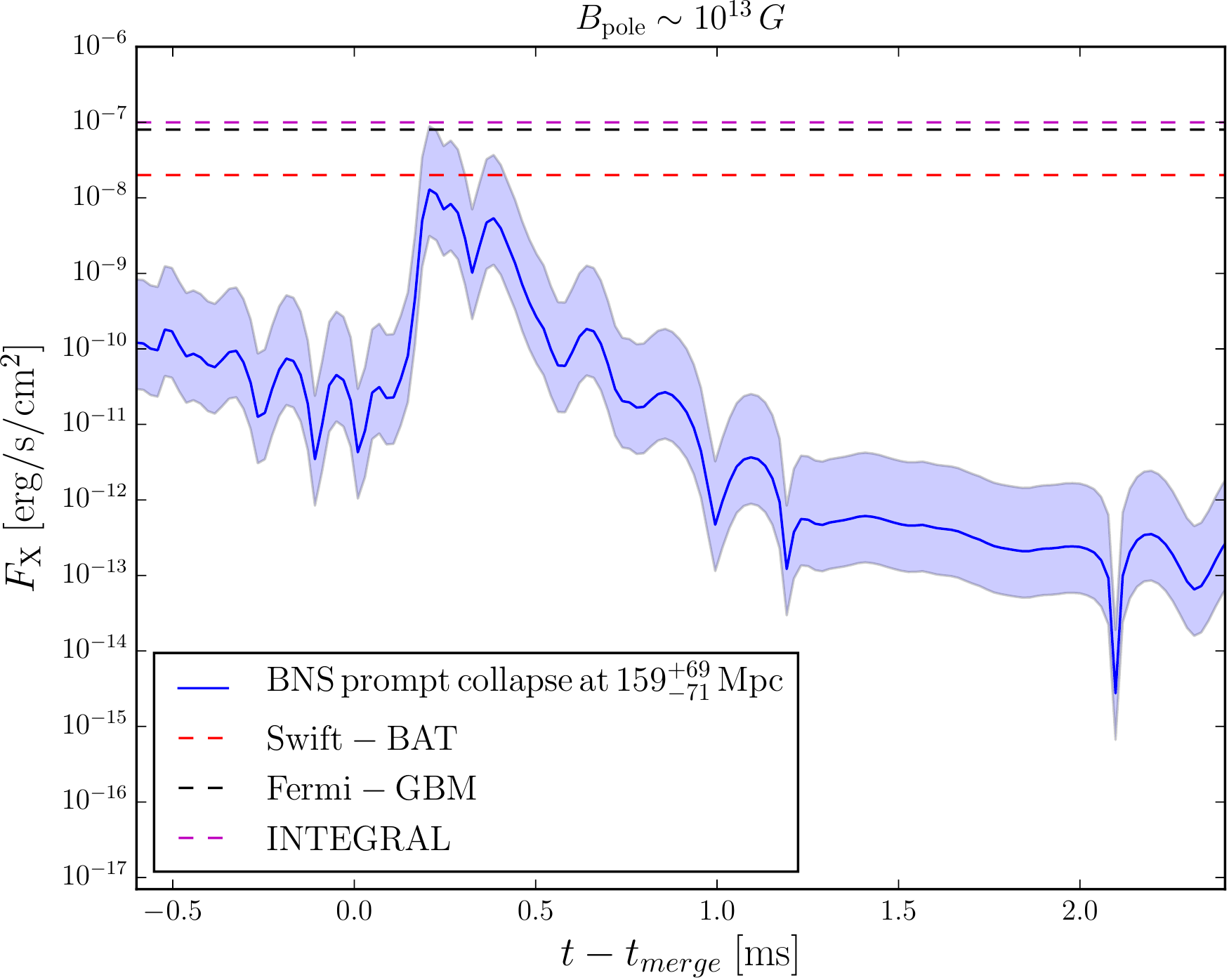}
	\end{center}
	\caption{Estimated flux from a BNS that promptly collapsed to a BH at 
	a distance of $159^{+69}_{-71}\, {\rm Mpc}$, with shaded blue the upper 
	and lower limits for the flux. The dashed lines correspond to the flux 
	limits of satellites that look blindly for burst-like events, 
	(dashed red)  \textit{Swift (BAT)} $15-100\,{\rm kev}$, (dashed black)
	\textit{Fermi (GBM)} $20\,{\rm kev}\,-40\,{\rm Mev}$ and 
	(dashed magenta)  \textit{INTEGRAL (IBAS)} $20-200\,{\rm kev}$, where 
	we have converted their photon rate in their respective energy 
	band, into observed flux.}
	\label{fig:fluxlim}
\end{figure}

At the time of merger the compression of the two magnetospheres leads to a strong
peak in the luminosity peak, see fig. \ref{fig:poynt} which 
is shown in retarded time to account for the time delay to reach 
the detector. At the time the actual burst is detected, the radiation 
from the initial transient, the "junk radiation" depicted at 
$0\, {\rm ms}$, has decreased almost two orders of magnitude, this gives 
us confidence that the main luminosity peak is not affected by this.  
It is important to note that the luminosity is increased in the anti-aligned case 
compared to the aligned case, whereas the unequal magnetic field
configuration lies in between the other two.
We speculate that this may be due to the alternating magnetic 
polarity close to the surface of the two stars. A similar behavior was 
also found in the inspiral of two neutron stars with their magnetospheres 
evolved in a highly conducting medium \citep{Palenzuela2013a}. 
The EM luminosity reported here is  expected to be similar with the 
one from a prompt collapse after  a BNS merger, however two points need 
to be made here, firstly the magnitude of the expected luminosity 
compared with what we find may be overestimated, since dissipation 
from magnetic reconnection via current sheets may reduce the magnetic 
energy, second this dissipated magnetic energy in current sheets can be
a source of high energy radiation potentially observable when a BNS 
merger remnant undergoes a prompt collapse to a BH, 
this is something that we further discuss in fig. \ref{fig:fluxlim}.

In all three cases after the merger the signal and the EM 
fields decay rapidly.
This is not surprising since a non-spinning BH is formed, which
cannot support a stationary magnetic field configuration, unlike for
spinning BHs where a Kerr-Newman BH can be formed
\citep{Nathanail2017}. 
At merger a transient is formed, which depends on the precise distortion of
the magnetospheres and the space time. After a BH has formed it
will ring down to a stationary Schwarzschild solution and consequently
the imprint of the quasi-normal modes of the BH are also found in
the decaying magnetic field, much like in the gravitational wave signal.
A similar feature has been observed and studied in previous
simulations of collapsing magnetized neutron stars with electrovacuum
magnetospheres \citep{Most2017,Dionysopoulou:2012pp,Baumgarte02b}.
 
Next, we compute the overall emitted EM energy 
during these events. The goal is to provide an analytic 
description of the radiated energy in terms of magnetic field 
strength and the corresponding efficiency in terms of 
the initial magnetic field orientation, in analogy to the case
of single star collapse \citep{Most2017}. Accordingly, we
compute the EM energy as:
\begin{equation}
E_{_{\rm EM}} := \int L_{_{\rm EM}}(t)\,dt \,,
\end{equation}
and report it in Table \ref{tab:initial} for all different intial
configurations. Similar to what was found for collapsing isolated neutron
stars \citep{Most2017}, we observe a perfect $B^2$ dependency of the
energy as expected on dimensional grounds (fig. \ref{fig:poynt}, upper 
panel). 

\begin{figure}
	\begin{center}	
	\includegraphics[width=1.0\columnwidth]{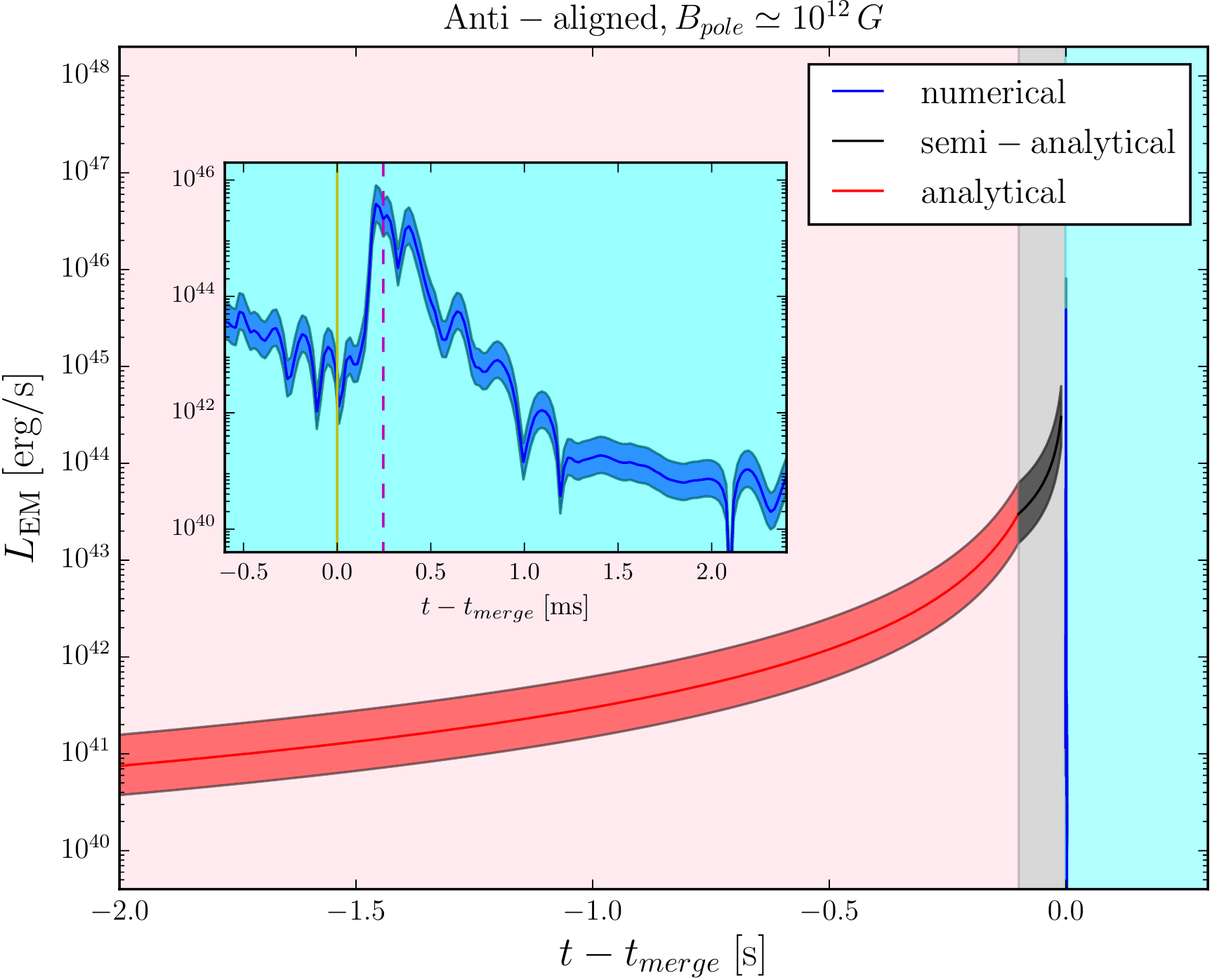}
	\end{center}
	\caption{ 
	  EM luminosity for the \texttt{Anti-Al$_2$} model. Zoom-in panel: 
	the numerical lightcurve, the solid 
	yellow line 	represents the time of merger and the dashed magenta line 
	the time when the apparent horizon is found for the first time. 
	Big panel:the precursor signal is reproduced by Eq. \eqref{eq:analytic} 
	till one hundred milliseconds before merger, then for the last hundred
	 milliseconds the scaling $L_{_{\rm EM}}\propto \Omega^{3/2}$, 
	 where $\Omega$ is the orbital frequency of the binary, is used.
	After merger the signal is the output of our numerical simulation. 
	It mimics the expected EM signature from a quasi-circular binary that 
	undergoes a prompt collapse in $0.5 {\rm ms }$.}
	\label{fig:joint_poynt}
\end{figure}
The reason to run all 
these models \texttt{$Al_1$}, \texttt{$Al_2$},\texttt{$Al_3$} and 
\texttt{$Al_4$}, is to test the dependence of the luminosity
with the initial magnetic field strength 
in the range of $\sim 10^{11} - 10^{14} \, {\rm G}$ (for the initial 
magnetic field at the pole), and then use this scaling to extrapolate 
also for the other models in the same range\footnote{As discussed in Sec. 
\ref{sec:mis}, the initial plasma parameter $\beta$ inside the stars is 
different for different initial magnetic field strength of the neutron 
stars, so the magnetic field  scaling needs to be tested.}. We 
find an almost perfect $\propto B^2$ scaling that holds in this range, 
with even the smallest features on the lightcurve being identical in all 
four cases. This gives us confidence that the  results 
of all other models ca be safely extrapolated 
in this range of magnetic field strengths.
In the embedded plot we present the comparison with the high resolution 
run for model \texttt{Anti-Al$_1$}, the main peak of the lightcurve, that 
carries most of the energy, is in good  agreement both in 
terms of energy and the time when the peak occurs.

In order to quantify the main differences of the emission efficiency 
due to the various initial magnetic field configurations, we 
estimate the available power in the two magnetospheres using 
a modified version of the phenomenological expression proposed by \cite{Falcke2013}. 
\begin{equation}
\label{eq:powerMS}
P_{_{\rm MS}} \simeq 16.8 \times 10^{44} \;
\eta_{_{\rm B}} \, t_{\rm ms}^{-1}\, b_{12}^2 
\ \ {\rm erg\ s}^{-1}\,,
\end{equation}
where $\eta_{_{\rm B}}$ is the EM efficiency, 
the fraction of the dissipated magnetic 
energy, $\Delta t = t_{\rm ms}\,1 {\rm ms}$ is the duration of the 
peak in the luminosity curve, while $b_{12}$ is the polar 
magnetic field of the star in units of $10^{12}\,{\rm G}$. 
Notice, that here we report the polar value and not the
maximum magnetic field, in order to allow for straight 
comparisons with the respective magnetic efficiency from 
single collapsing NS.

In all models the burst-like high luminosity peak is of the order of 
 one millisecond, thus  $\Delta t_{_{\rm EM}}/{\rm ms} = 1 = t_{\rm ms}$. 
The emitted energy can be expressed as:
\begin{equation}
\label{eq:scaled_EEM}
E_{_{\rm EM}} = P_{_{\rm MS}}\, \Delta t_{_{\rm EM}}
\simeq 16.8 \times 10^{41} \;
\eta_{_{\rm B}} \, b_{12}^2 
\ \ {\rm erg}\,.
\end{equation}                          
The duration of these burst-like peaks in luminosity is dictated 
by the matter dynamics and the almost immediate  collapse to a BH, 
this is the reason why the simulations described here, can be 
regarded as a toy model for BNS mergers that promptly collapse to a
BH, since the head-on collision is not related to any of the features 
on the  luminosity lightcurve.
From the radiated energy for the different models in Table 
\ref{tab:initial} and expression \ref{eq:scaled_EEM}, we can deduce 
the magnetic efficiency $\eta_{_{\rm B}}$. For the three models 
with the different magnetic field initial configuration we find: 
for \texttt{Anti-al$_1$}  $\eta_{_{\rm B}}\simeq 12 \%$, 
for \texttt{unequal-B} $\eta_{_{\rm B}}\simeq 3.7 \%$ and 
\texttt{Al$_1$} $\eta_{_{\rm B}}\simeq 2 \%$. 
Several important remarks can be made here. As we followed 
the evolution of the \texttt{Al$_1$} model in Sec. \ref{sec:nsaid}, 
where both magnetic dipoles are initially aligned, we argued that 
after merger the evolution closely mimics that
of a collapsing magnetised NS. It was found that the mean magnetic 
efficiency from such a collapse is $\eta_{_{\rm B}}\simeq 2 \%$
\citep{Most2017}, which is exactly what we get for the aligned case. In the 
case with unequal magnetic field, the efficiency is almost twice 
higher and in the most efficient magnetic field configuration, 
the anti-aligned dipoles, the efficiency is six times higher 
than the aligned one.

The computed estimates for the radiated energy and 
luminosity, together with the luminosity curves of fig. \ref{fig:poynt}, 
point that the EM energy radiated in such events can potentially be 
 compared  with  the phenomenology 
of FRBs  (for a review \cite{RaneLorimer2017}). Although the systems
studied in this work are highly idealised and merely serve as a toy model
to investigate a prompt collapse scenario, we can still draw qualitative
conclusions relevant for realistic BNS configurations. 
In particular BNS systems with very high masses are known to undergo 
a prompt collapse \citep{Hotokezaka2011, Bauswein2013,Ruiz2017a,Koeppel2019}, 
where a light disk is formed \citep{Nathanail2018,Paschalidis2018}. 
The lifetime of this disk is found to be of the order of milliseconds 
\citep{Ruiz2017a}. Thus, the disk will not be able to keep the magnetic
field from dissipating and a large fraction 
 of the  magnetic energy will 
be radiated away in a similar manner to the one we have described
in this study.


%

The expected lightcurve from a quasi-circular binary of NS 
that undergoes a prompt collapse will consist also with a precursor 
signal. In order to produce the lightcurve we use an analytic formula
for the expected precursor signal in the case where only one NS is 
magnetised, this case is resembled in our study with the 
\texttt{Anti-Al$_2$} model. The luminosity of such a binary  is:
\begin{equation}
\label{eq:analytic}
L_{_{\rm EM}} 
\simeq 3 \times 10^{41}  \;
 \, b_{12}^2 \, (-t)^{-7/4}
\ \ {\rm erg s^{-1}}\,.
\end{equation}                          
where $t$  the time to merger  is measured in seconds and $b_{12}$ as
before \citep{Hansen2001,Lyutikov:2011c,Lyutikov2018}. 
Equation \eqref{eq:analytic} would overestimate the luminosity  as 
the system approaches merger, diverging at the merger point. 
Thus, we use the results from \citep{Palenzuela2013b} 
 where they report  the luminosity prior to merger, for a binary 
 neutron star system where the magnetic dipoles are anti-aligned,
 scales like: 
$L_{_{\rm EM}}\propto \Omega^{3/2}$, where $\Omega$ is the orbital frequency 
of the binary. This is smoothly joint with the previous analytic expression 
for the last one hundred milliseconds before merger.
In fig. \ref{fig:joint_poynt} we combine these formulas with 
our numerical results to represent the expected EM luminosity for 
the case of quasi-circular binary that undergoes a prompt collapse 
to a BH in the first $0.5 {\rm ms}$.

The latest detection of GW from a BNS merger, GW190425, unveiled a 
binary with a total 
mass of $3.4^{+0.3}_{-0.1}\, M_{\odot}$, which is rather unusual with regard to the 
binaries observed in our galaxy, and is consistent with a distance 
of  $159^{+69}_{-71}\, {\rm Mpc}$ \citep{Abbott2020}. 
The current status about 
a binary of this total mass, expects a prompt collapse to a BH
\citep{Bauswein2013,Hotokezaka2011,Koeppel2019,Agathos2019}.
We will use the toy model presented in this study in order to set limits 
on the flux expected from a BNS merging at the  distance  of GW190425. 
We assume that the magnetic loops radiated in  our simulations dissipate 
magnetic energy through magnetic reconnection and a fraction of this 
 is emitted as high-energy photons at the energy range of satellites 
that search for triggers of burst -like events, such as
\textit{Swift (BAT)}, \textit{Fermi (GBM)} and \textit{INTEGRAL}. 
The estimated flux is:
\begin{align}
 F_X = f\frac{L}{4\pi d^2_L} \, .
 \label{eqn:flux}
\end{align}
where $L$ is the luminosity coming from the simulation, $d_L$ 
is the luminosity distance and $f$ the fraction of the luminosity 
that dissipates in high-energy photons, which we assume to be 
$f=10\%$, since similar values are found from kinetic simulations 
for dissipation through reconnection in pulsar magnetospheres
\citep{Philippov2014,Cerutti2015,Brambilla2018,Crinquand2019}.

The expected flux is shown in fig. \ref{fig:fluxlim},  for 
a detection of a BNS that supposedly promptly collapsed to a BH, 
at a  distance of $159^{+69}_{-71}\, {\rm Mpc}$,
where the upper and lower limits on the distance correspond to the upper 
and lower flux limits shaded in blue, also shown are the limits 
of the high-energy detectors from  \textit{Swift (BAT)}\cite{Barthelmy2005b}, 
\textit{Fermi (GBM)} \cite{Meegan2009}
and \textit{INTEGRAL (IBAS)} \cite{Mereghetii2003}. 
From the analysis provided by our modeling it is evident that 
a detection would not be possible, since the distance to the 
source is placing the expected flux exactly below the limits of 
the detectors.
As a last remark, we should point out that in terms of radio, 
such a BNS is expected to have a radio luminosity with an efficiency 
of $10^{-3}-10^{-7}$ \citep{Szary2014}, 
since all observed pulsars have a similar radio 
efficiency when comparing the expected radiation from the magnetosphere
with the observed one, and thus is extremely far to discuss any 
possibility for detection.

\section{Conclusions}
\label{sec:con}
In the dawn of the multi-messenger era, where 
GW and EM radiation can be observed simultaneously 
from the merger of binary NS systems, any attempt of 
clarifying such physical picture is invaluable. 
In this respect, the understanding  of  the dissipation 
of the magnetic energy of the two NS, when the system quickly 
undergoes a collapse to a BH, with no disk (or a minor disk) 
surrounding it, is of great importance. In this study we have 
followed a more idealized case of the head on collision of 
two NSs, in order to study the radiated EM energy 
produced during such collisions. 

We have presented a comprehensive study of the head on collision 
of two magnetised NS. We have performed simulations with different 
initial magnetic field configurations and strengths. We have further 
deduced the efficiency of dissipating the available energy 
stored in both magnetospheres. We have shown that in the case of two 
anti-ligned magnetic dipoles the EM luminosity has an excess 
of two orders of magnitude, compared with the respective one 
coming from the merger of two stars with aligned dipoles, this result 
has been reported in studies of precursor signals 
\citep{Palenzuela2013b}. 
Our results were closely compared 
with the respective collapse of a magnetised NS.  
The EM luminosity is dictated by the collapse timescale and the magnetic 
energy stored in the magnetosphere, and not by the head-on collision 
assumption, this is the reason why we can use this study as a toy 
model and a conservative estimate for the EM output of 
a BNS prompt collapse to a BH, 
since the collapse timescale in this case is close 
to $\sim 1\, {\rm ms}$.

Lastly, we discussed the resemblance of such EM signals of 
millisecond duration with the phenomenology of FRBs. 
We further discussed the possibility that such signals may 
originate from BNS mergers that 
undergo prompt collapse, and provided the expected 
high-energy flux limits, 
according to our analysis with respect to the GW detection 
GW190425, concluding that it would not be possible to trigger 
any of the satellite detectors considered.


\section*{Acknowledgements}

The author is thankful to L. Rezzolla and E. Most for useful discussions.
AN is partially supported by an Alexander von Humboldt Fellowship. Partial 
support comes  from ``NewCompStar'', COST Action
MP1304.
The simulations were performed on SuperMUC at
LRZ-Munich, on LOEWE at CSC-Frankfurt and on Hazelhen at HLRS in
Stuttgart.







\bibliographystyle{mnras}
\bibliography{aeireferences}

\begin{thebibliography}{}
\makeatletter
\relax
\def\mn@urlcharsother{\let\do\@makeother \do\$\do\&\do\#\do\^\do\_\do\%\do\~}
\def\mn@doi{\begingroup\mn@urlcharsother \@ifnextchar [ {\mn@doi@}
  {\mn@doi@[]}}
\def\mn@doi@[#1]#2{\def\@tempa{#1}\ifx\@tempa\@empty \href
  {http://dx.doi.org/#2} {doi:#2}\else \href {http://dx.doi.org/#2} {#1}\fi
  \endgroup}
\def\mn@eprint#1#2{\mn@eprint@#1:#2::\@nil}
\def\mn@eprint@arXiv#1{\href {http://arxiv.org/abs/#1} {{\tt arXiv:#1}}}
\def\mn@eprint@dblp#1{\href {http://dblp.uni-trier.de/rec/bibtex/#1.xml}
  {dblp:#1}}
\def\mn@eprint@#1:#2:#3:#4\@nil{\def\@tempa {#1}\def\@tempb {#2}\def\@tempc
  {#3}\ifx \@tempc \@empty \let \@tempc \@tempb \let \@tempb \@tempa \fi \ifx
  \@tempb \@empty \def\@tempb {arXiv}\fi \@ifundefined
  {mn@eprint@\@tempb}{\@tempb:\@tempc}{\expandafter \expandafter \csname
  mn@eprint@\@tempb\endcsname \expandafter{\@tempc}}}

\bibitem[\protect\citeauthoryear{Abbott et~al.}{Abbott
  et~al.}{2017}]{Abbott2017d_etal}
Abbott B.~P.,  et~al., 2017, Astrophys. J. Lett., 848, L13

\bibitem[\protect\citeauthoryear{{Agathos}, {Zappa}, {Bernuzzi}, {Perego},
  {Breschi}  \& {Radice}}{{Agathos} et~al.}{2019}]{Agathos2019}
{Agathos} M.,  {Zappa} F.,  {Bernuzzi} S.,  {Perego} A.,  {Breschi} M.,
  {Radice} D.,  2019, arXiv e-prints, \href
  {https://ui.adsabs.harvard.edu/abs/2019arXiv190805442A} {p. arXiv:1908.05442}

\bibitem[\protect\citeauthoryear{{Alic}, {Bona-Casas}, {Bona}, {Rezzolla}  \&
  {Palenzuela}}{{Alic} et~al.}{2012}]{Alic:2011a}
{Alic} D.,  {Bona-Casas} C.,  {Bona} C.,  {Rezzolla} L.,   {Palenzuela} C.,
  2012, \mn@doi [Phys. Rev. D] {10.1103/PhysRevD.85.064040}, \href
  {http://adsabs.harvard.edu/abs/2012PhRvD..85f4040A} {85, 064040}

\bibitem[\protect\citeauthoryear{{Alic}, {Kastaun}  \& {Rezzolla}}{{Alic}
  et~al.}{2013}]{Alic2013}
{Alic} D.,  {Kastaun} W.,   {Rezzolla} L.,  2013, \mn@doi [Phys. Rev. D]
  {10.1103/PhysRevD.88.064049}, \href
  {http://adsabs.harvard.edu/abs/2013PhRvD..88f4049A} {88, 064049}

\bibitem[\protect\citeauthoryear{Baiotti \& Rezzolla}{Baiotti \&
  Rezzolla}{2017}]{Baiotti2016}
Baiotti L.,  Rezzolla L.,  2017, \mn@doi [Rept. Prog. Phys.]
  {10.1088/1361-6633/aa67bb}, 80, 096901

\bibitem[\protect\citeauthoryear{{Baiotti}, {Giacomazzo}  \&
  {Rezzolla}}{{Baiotti} et~al.}{2008}]{Baiotti08}
{Baiotti} L.,  {Giacomazzo} B.,   {Rezzolla} L.,  2008, \mn@doi [Phys. Rev. D]
  {10.1103/PhysRevD.78.084033}, \href
  {http://adsabs.harvard.edu/abs/2008PhRvD..78h4033B} {78, 084033}

\bibitem[\protect\citeauthoryear{{Barthelmy} et~al.,}{{Barthelmy}
  et~al.}{2005}]{Barthelmy2005b}
{Barthelmy} S.~D.,  et~al., 2005, \mn@doi [Space Science Reviews]
  {10.1007/s11214-005-5096-3}, \href
  {https://ui.adsabs.harvard.edu/abs/2005SSRv..120..143B} {120, 143}

\bibitem[\protect\citeauthoryear{Baumgarte \& Shapiro}{Baumgarte \&
  Shapiro}{2002}]{Baumgarte02b}
Baumgarte T.~W.,  Shapiro S.~L.,  2002

\bibitem[\protect\citeauthoryear{{Bauswein}, {Baumgarte}  \&
  {Janka}}{{Bauswein} et~al.}{2013}]{Bauswein2013}
{Bauswein} A.,  {Baumgarte} T.~W.,   {Janka} H.-T.,  2013, \mn@doi [Phys. Rev.
  Lett.] {10.1103/PhysRevLett.111.131101}, \href
  {http://adsabs.harvard.edu/abs/2013PhRvL.111m1101B} {111, 131101}

\bibitem[\protect\citeauthoryear{{Bovard}, {Martin}, {Guercilena}, {Arcones},
  {Rezzolla}  \& {Korobkin}}{{Bovard} et~al.}{2017}]{Bovard2017}
{Bovard} L.,  {Martin} D.,  {Guercilena} F.,  {Arcones} A.,  {Rezzolla} L.,
  {Korobkin} O.,  2017, Phys. Rev. D, \href
  {http://adsabs.harvard.edu/abs/2017arXiv170909630B} {96, 124005}

\bibitem[\protect\citeauthoryear{{Brambilla}, {Kalapotharakos}, {Timokhin},
  {Harding}  \& {Kazanas}}{{Brambilla} et~al.}{2018}]{Brambilla2018}
{Brambilla} G.,  {Kalapotharakos} C.,  {Timokhin} A.~N.,  {Harding} A.~K.,
  {Kazanas} D.,  2018, \mn@doi [Astrophys. J.] {10.3847/1538-4357/aab3e1},
  \href {https://ui.adsabs.harvard.edu/abs/2018ApJ...858...81B} {858, 81}

\bibitem[\protect\citeauthoryear{{Bucciantini} \& {Del Zanna}}{{Bucciantini} \&
  {Del Zanna}}{2013}]{Bucciantini2012a}
{Bucciantini} N.,  {Del Zanna} L.,  2013, \mn@doi [Mon. Not. R. Astron. Soc.]
  {10.1093/mnras/sts005}, \href
  {http://adsabs.harvard.edu/abs/2013MNRAS.428...71B} {428, 71}

\bibitem[\protect\citeauthoryear{{Cerutti}, {Philippov}, {Parfrey}  \&
  {Spitkovsky}}{{Cerutti} et~al.}{2015}]{Cerutti2015}
{Cerutti} B.,  {Philippov} A.,  {Parfrey} K.,   {Spitkovsky} A.,  2015, \mn@doi
  [Mon. Not. R. Astron. Soc.] {10.1093/mnras/stv042}, \href
  {https://ui.adsabs.harvard.edu/abs/2015MNRAS.448..606C} {448, 606}

\bibitem[\protect\citeauthoryear{{Ciolfi}, {Kastaun}, {Giacomazzo}, {Endrizzi},
  {Siegel}  \& {Perna}}{{Ciolfi} et~al.}{2017}]{Ciolfi2017}
{Ciolfi} R.,  {Kastaun} W.,  {Giacomazzo} B.,  {Endrizzi} A.,  {Siegel} D.~M.,
   {Perna} R.,  2017, \mn@doi [Phys. Rev. D] {10.1103/PhysRevD.95.063016},
  \href {http://adsabs.harvard.edu/abs/2017PhRvD..95f3016C} {95, 063016}

\bibitem[\protect\citeauthoryear{{Colella} \& {Sekora}}{{Colella} \&
  {Sekora}}{2008}]{Colella2008}
{Colella} P.,  {Sekora} M.~D.,  2008, \mn@doi [Journal of Computational
  Physics] {10.1016/j.jcp.2008.03.034}, \href
  {http://adsabs.harvard.edu/abs/2008JCoPh.227.7069C} {227, 7069}

\bibitem[\protect\citeauthoryear{{Crinquand}, {Cerutti}  \&
  {Dubus}}{{Crinquand} et~al.}{2019}]{Crinquand2019}
{Crinquand} B.,  {Cerutti} B.,   {Dubus} G.,  2019, \mn@doi [Astron.
  Astrophys.] {10.1051/0004-6361/201834610}, \href
  {https://ui.adsabs.harvard.edu/abs/2019A&A...622A.161C} {622, A161}

\bibitem[\protect\citeauthoryear{{Dietrich} \& {Ujevic}}{{Dietrich} \&
  {Ujevic}}{2017}]{Dietrich2016}
{Dietrich} T.,  {Ujevic} M.,  2017, \mn@doi [Classical and Quantum Gravity]
  {10.1088/1361-6382/aa6bb0}, \href
  {http://adsabs.harvard.edu/abs/2017CQGra..34j5014D} {34, 105014}

\bibitem[\protect\citeauthoryear{{Dietrich}, {Ujevic}, {Tichy}, {Bernuzzi}  \&
  {Br{\"u}gmann}}{{Dietrich} et~al.}{2017a}]{Dietrich2017}
{Dietrich} T.,  {Ujevic} M.,  {Tichy} W.,  {Bernuzzi} S.,   {Br{\"u}gmann} B.,
  2017a, \mn@doi [Phys. Rev. D] {10.1103/PhysRevD.95.024029}, \href
  {http://adsabs.harvard.edu/abs/2017PhRvD..95b4029D} {95, 024029}

\bibitem[\protect\citeauthoryear{{Dietrich}, {Bernuzzi}  \& {Tichy}}{{Dietrich}
  et~al.}{2017b}]{Dietrich2017b}
{Dietrich} T.,  {Bernuzzi} S.,   {Tichy} W.,  2017b, \mn@doi [Phys. Rev. D]
  {10.1103/PhysRevD.96.121501}, \href
  {http://adsabs.harvard.edu/abs/2017PhRvD..96l1501D} {96, 121501}

\bibitem[\protect\citeauthoryear{{Dionysopoulou}, {Alic}, {Palenzuela},
  {Rezzolla}  \& {Giacomazzo}}{{Dionysopoulou}
  et~al.}{2013}]{Dionysopoulou:2012pp}
{Dionysopoulou} K.,  {Alic} D.,  {Palenzuela} C.,  {Rezzolla} L.,
  {Giacomazzo} B.,  2013, \mn@doi [Phys. Rev. D] {10.1103/PhysRevD.88.044020},
  \href {http://adsabs.harvard.edu/abs/2013PhRvD..88d4020D} {88, 044020}

\bibitem[\protect\citeauthoryear{{Dionysopoulou}, {Alic}  \&
  {Rezzolla}}{{Dionysopoulou} et~al.}{2015}]{Dionysopoulou2015}
{Dionysopoulou} K.,  {Alic} D.,   {Rezzolla} L.,  2015, \mn@doi [Phys. Rev. D]
  {10.1103/PhysRevD.92.084064}, \href
  {http://adsabs.harvard.edu/abs/2015PhRvD..92h4064D} {92, 084064}

\bibitem[\protect\citeauthoryear{{East} \& {Pretorius}}{{East} \&
  {Pretorius}}{2013}]{East2012}
{East} W.~E.,  {Pretorius} F.,  2013, \mn@doi [Phys. Rev. Lett.]
  {10.1103/PhysRevLett.110.101101}, \href
  {http://adsabs.harvard.edu/abs/2013PhRvL.110j1101E} {110, 101101}

\bibitem[\protect\citeauthoryear{{Falcke} \& {Rezzolla}}{{Falcke} \&
  {Rezzolla}}{2014}]{Falcke2013}
{Falcke} H.,  {Rezzolla} L.,  2014, \mn@doi [Astron. Astrophys.]
  {10.1051/0004-6361/201321996}, \href
  {http://adsabs.harvard.edu/abs/2014A%26A...562A.137F} {562, A137}

\bibitem[\protect\citeauthoryear{{Fern{\'a}ndez}, {Tchekhovskoy}, {Quataert},
  {Foucart}  \& {Kasen}}{{Fern{\'a}ndez} et~al.}{2018}]{Fernandez2018}
{Fern{\'a}ndez} R.,  {Tchekhovskoy} A.,  {Quataert} E.,  {Foucart} F.,
  {Kasen} D.,  2018, \mn@doi [Mon. Not. R. Astron. Soc.]
  {10.1093/mnras/sty2932}, \href
  {http://adsabs.harvard.edu/abs/2018MNRAS.tmp.2798F} {}

\bibitem[\protect\citeauthoryear{{Foucart}, {O'Connor}, {Roberts}, {Kidder},
  {Pfeiffer}  \& {Scheel}}{{Foucart} et~al.}{2016}]{Foucart2016a}
{Foucart} F.,  {O'Connor} E.,  {Roberts} L.,  {Kidder} L.~E.,  {Pfeiffer}
  H.~P.,   {Scheel} M.~A.,  2016, \mn@doi [Phys. Rev. D]
  {10.1103/PhysRevD.94.123016}, \href
  {http://adsabs.harvard.edu/abs/2016PhRvD..94l3016F} {94, 123016}

\bibitem[\protect\citeauthoryear{{Fujibayashi}, {Sekiguchi}, {Kiuchi}  \&
  {Shibata}}{{Fujibayashi} et~al.}{2017}]{Fujibayashi2017}
{Fujibayashi} S.,  {Sekiguchi} Y.,  {Kiuchi} K.,   {Shibata} M.,  2017, \mn@doi
  [Astrophys. J.] {10.3847/1538-4357/aa8039}, \href
  {http://adsabs.harvard.edu/abs/2017ApJ...846..114F} {846, 114}

\bibitem[\protect\citeauthoryear{{Fujibayashi}, {Kiuchi}, {Nishimura},
  {Sekiguchi}  \& {Shibata}}{{Fujibayashi} et~al.}{2018}]{Fujibayashi2017b}
{Fujibayashi} S.,  {Kiuchi} K.,  {Nishimura} N.,  {Sekiguchi} Y.,   {Shibata}
  M.,  2018, \mn@doi [Astrophys. J.] {10.3847/1538-4357/aabafd}, \href
  {http://adsabs.harvard.edu/abs/2018ApJ...860...64F} {860, 64}

\bibitem[\protect\citeauthoryear{Gundlach, Martin-Garcia, Calabrese  \&
  Hinder}{Gundlach et~al.}{2005}]{Gundlach2005:constraint-damping}
Gundlach C.,  Martin-Garcia J.~M.,  Calabrese G.,   Hinder I.,  2005, \mn@doi
  [Class. Quantum Grav.] {10.1088/0264-9381/22/17/025}, 22, 3767

\bibitem[\protect\citeauthoryear{{Hanauske}, {Takami}, {Bovard}, {Rezzolla},
  {Font}, {Galeazzi}  \& {St{\"o}cker}}{{Hanauske} et~al.}{2017}]{Hanauske2016}
{Hanauske} M.,  {Takami} K.,  {Bovard} L.,  {Rezzolla} L.,  {Font} J.~A.,
  {Galeazzi} F.,   {St{\"o}cker} H.,  2017, \mn@doi [Phys. Rev. D]
  {10.1103/PhysRevD.96.043004}, \href
  {http://adsabs.harvard.edu/abs/2017PhRvD..96d3004H} {96, 043004}

\bibitem[\protect\citeauthoryear{Hansen \& Lyutikov}{Hansen \&
  Lyutikov}{2001}]{Hansen2001}
Hansen B. M.~S.,  Lyutikov M.,  2001, \mn@doi [Mon. Not. R. Astron. Soc.]
  {10.1046/j.1365-8711.2001.04103.x}, 322, 695

\bibitem[\protect\citeauthoryear{{Harutyunyan}, {Nathanail}, {Rezzolla}  \&
  {Sedrakian}}{{Harutyunyan} et~al.}{2018}]{Harutyunyan2018}
{Harutyunyan} A.,  {Nathanail} A.,  {Rezzolla} L.,   {Sedrakian} A.,  2018,
  \mn@doi [European Physical Journal A] {10.1140/epja/i2018-12624-1}, \href
  {https://ui.adsabs.harvard.edu/abs/2018EPJA...54..191H} {54, 191}

\bibitem[\protect\citeauthoryear{{Hotokezaka} \& {Piran}}{{Hotokezaka} \&
  {Piran}}{2015}]{Hotokezaka2015MNRAS}
{Hotokezaka} K.,  {Piran} T.,  2015, \mn@doi [Mon. Not. R. Astron. Soc.]
  {10.1093/mnras/stv620}, \href
  {http://adsabs.harvard.edu/abs/2015MNRAS.450.1430H} {450, 1430}

\bibitem[\protect\citeauthoryear{{Hotokezaka}, {Kyutoku}, {Okawa}, {Shibata}
  \& {Kiuchi}}{{Hotokezaka} et~al.}{2011}]{Hotokezaka2011}
{Hotokezaka} K.,  {Kyutoku} K.,  {Okawa} H.,  {Shibata} M.,   {Kiuchi} K.,
  2011, \mn@doi [Phys. Rev. D] {10.1103/PhysRevD.83.124008}, \href
  {http://adsabs.harvard.edu/abs/2011PhRvD..83l4008H} {83, 124008}

\bibitem[\protect\citeauthoryear{{Hotokezaka}, {Piran}  \& {Paul}}{{Hotokezaka}
  et~al.}{2017}]{Hotokezaka:2015b}
{Hotokezaka} K.,  {Piran} T.,   {Paul} M.,  2017, in {Kubono} S.,  {Kajino} T.,
   {Nishimura} S.,  {Isobe} T.,  {Nagataki} S.,  {Shima} T.,   {Takeda} Y.,
  eds, 14th International Symposium on Nuclei in the Cosmos (NIC2016). p.
  010608 (\mn@eprint {arXiv} {1510.00711}), \mn@doi{10.7566/JPSCP.14.010608}

\bibitem[\protect\citeauthoryear{{Kastaun}, {Galeazzi}, {Alic}, {Rezzolla}  \&
  {Font}}{{Kastaun} et~al.}{2013}]{Kastaun2013}
{Kastaun} W.,  {Galeazzi} F.,  {Alic} D.,  {Rezzolla} L.,   {Font} J.~A.,
  2013, \mn@doi [Phys. Rev. D] {10.1103/PhysRevD.88.021501}, \href
  {http://adsabs.harvard.edu/abs/2013PhRvD..88b1501K} {88, 021501}

\bibitem[\protect\citeauthoryear{{Kastaun}, {Ciolfi}, {Endrizzi}  \&
  {Giacomazzo}}{{Kastaun} et~al.}{2016a}]{Kastaun2016b}
{Kastaun} W.,  {Ciolfi} R.,  {Endrizzi} A.,   {Giacomazzo} B.,  2016a,
  preprint, \href {http://adsabs.harvard.edu/abs/2016arXiv161203671K} {}
  (\mn@eprint {arXiv} {1612.03671})

\bibitem[\protect\citeauthoryear{{Kastaun}, {Ciolfi}  \&
  {Giacomazzo}}{{Kastaun} et~al.}{2016b}]{Kastaun2016}
{Kastaun} W.,  {Ciolfi} R.,   {Giacomazzo} B.,  2016b, \mn@doi [Phys. Rev. D]
  {10.1103/PhysRevD.94.044060}, \href
  {http://adsabs.harvard.edu/abs/2016PhRvD..94d4060K} {94, 044060}

\bibitem[\protect\citeauthoryear{{Kellerman}, {Rezzolla}  \&
  {Radice}}{{Kellerman} et~al.}{2010}]{Kellermann:10}
{Kellerman} T.,  {Rezzolla} L.,   {Radice} D.,  2010, \mn@doi [Class. Quantum
  Grav.] {10.1088/0264-9381/27/23/235016}, \href
  {http://adsabs.harvard.edu/abs/2010CQGra..27w5016K} {27, 235016}

\bibitem[\protect\citeauthoryear{{Kiuchi}, {Kyutoku}, {Sekiguchi}, {Shibata}
  \& {Wada}}{{Kiuchi} et~al.}{2014}]{Kiuchi2014}
{Kiuchi} K.,  {Kyutoku} K.,  {Sekiguchi} Y.,  {Shibata} M.,   {Wada} T.,  2014,
  \mn@doi [Phys. Rev. D] {10.1103/PhysRevD.90.041502}, \href
  {http://adsabs.harvard.edu/abs/2014PhRvD..90d1502K} {90, 041502}

\bibitem[\protect\citeauthoryear{{Kiuchi}, {Sekiguchi}, {Kyutoku}, {Shibata},
  {Taniguchi}  \& {Wada}}{{Kiuchi} et~al.}{2015}]{Kiuchi2015}
{Kiuchi} K.,  {Sekiguchi} Y.,  {Kyutoku} K.,  {Shibata} M.,  {Taniguchi} K.,
  {Wada} T.,  2015, \mn@doi [Phys. Rev. D] {10.1103/PhysRevD.92.064034}, \href
  {http://adsabs.harvard.edu/abs/2015PhRvD..92f4034K} {92, 064034}

\bibitem[\protect\citeauthoryear{{K{\"o}ppel}, {Bovard}  \&
  {Rezzolla}}{{K{\"o}ppel} et~al.}{2019}]{Koeppel2019}
{K{\"o}ppel} S.,  {Bovard} L.,   {Rezzolla} L.,  2019, \mn@doi [Astrophys. J.
  Lett.] {10.3847/2041-8213/ab0210}, \href
  {https://ui.adsabs.harvard.edu/abs/2019ApJ...872L..16K} {872, L16}

\bibitem[\protect\citeauthoryear{{Lai}}{{Lai}}{2012}]{Lai2012}
{Lai} D.,  2012, \mn@doi [Astrophys. J. Lett.] {10.1088/2041-8205/757/1/L3},
  \href {http://adsabs.harvard.edu/abs/2012ApJ...757L...3L} {757, L3}

\bibitem[\protect\citeauthoryear{{Lasky}, {Haskell}, {Ravi}, {Howell}  \&
  {Coward}}{{Lasky} et~al.}{2014}]{Lasky2013}
{Lasky} P.~D.,  {Haskell} B.,  {Ravi} V.,  {Howell} E.~J.,   {Coward} D.~M.,
  2014, \mn@doi [Phys. Rev. D] {10.1103/PhysRevD.89.047302}, \href
  {http://adsabs.harvard.edu/abs/2014PhRvD..89d7302L} {89, 047302}

\bibitem[\protect\citeauthoryear{{Lehner}, {Liebling}, {Palenzuela},
  {Caballero}, {O'Connor}, {Anderson}  \& {Neilsen}}{{Lehner}
  et~al.}{2016}]{Lehner2016}
{Lehner} L.,  {Liebling} S.~L.,  {Palenzuela} C.,  {Caballero} O.~L.,
  {O'Connor} E.,  {Anderson} M.,   {Neilsen} D.,  2016, \mn@doi [Classical and
  Quantum Gravity] {10.1088/0264-9381/33/18/184002}, \href
  {http://adsabs.harvard.edu/abs/2016CQGra..33r4002L} {33, 184002}

\bibitem[\protect\citeauthoryear{Liu, Shapiro, Etienne  \& Taniguchi}{Liu
  et~al.}{2008}]{Liu:2008xy}
Liu Y.~T.,  Shapiro S.~L.,  Etienne Z.~B.,   Taniguchi K.,  2008, \mn@doi
  [Phys. Rev. D] {10.1103/PhysRevD.78.024012}, 78, 024012

\bibitem[\protect\citeauthoryear{{L{\"o}ffler} et~al.,}{{L{\"o}ffler}
  et~al.}{2012}]{loeffler_2011_et}
{L{\"o}ffler} F.,  et~al., 2012, \mn@doi [Class. Quantum Grav.]
  {10.1088/0264-9381/29/11/115001}, \href
  {http://adsabs.harvard.edu/abs/2012CQGra..29k5001L} {29, 115001}

\bibitem[\protect\citeauthoryear{{Lyutikov}}{{Lyutikov}}{2011}]{Lyutikov:2011c}
{Lyutikov} M.,  2011, \mn@doi [Phys. Rev. D] {10.1103/PhysRevD.83.124035},
  \href {http://adsabs.harvard.edu/abs/2011PhRvD..83l4035L} {83, 124035}

\bibitem[\protect\citeauthoryear{{Lyutikov}}{{Lyutikov}}{2018}]{Lyutikov2018}
{Lyutikov} M.,  2018, preprint, \href
  {http://adsabs.harvard.edu/abs/2018arXiv180910478L} {} (\mn@eprint {arXiv}
  {1809.10478})

\bibitem[\protect\citeauthoryear{{Margalit} \& {Metzger}}{{Margalit} \&
  {Metzger}}{2019}]{Margalit2019}
{Margalit} B.,  {Metzger} B.~D.,  2019, \mn@doi [Astrophys. J. Lett.]
  {10.3847/2041-8213/ab2ae2}, \href
  {https://ui.adsabs.harvard.edu/abs/2019ApJ...880L..15M} {880, L15}

\bibitem[\protect\citeauthoryear{{Meegan} et~al.,}{{Meegan}
  et~al.}{2009}]{Meegan2009}
{Meegan} C.,  et~al., 2009, \mn@doi [Astrophys. J.]
  {10.1088/0004-637X/702/1/791}, \href
  {https://ui.adsabs.harvard.edu/abs/2009ApJ...702..791M} {702, 791}

\bibitem[\protect\citeauthoryear{{Mereghetti}, {G{\"o}tz}, {Borkowski},
  {Walter}  \& {Pedersen}}{{Mereghetti} et~al.}{2003}]{Mereghetii2003}
{Mereghetti} S.,  {G{\"o}tz} D.,  {Borkowski} J.,  {Walter} R.,   {Pedersen}
  H.,  2003, \mn@doi [Astron. Astrophys.] {10.1051/0004-6361:20031289}, \href
  {https://ui.adsabs.harvard.edu/abs/2003A&A...411L.291M} {411, L291}

\bibitem[\protect\citeauthoryear{{Metzger} \& {Zivancev}}{{Metzger} \&
  {Zivancev}}{2016}]{Metzger2016}
{Metzger} B.~D.,  {Zivancev} C.,  2016, \mn@doi [Mon. Not. R. Astron. Soc.]
  {10.1093/mnras/stw1800}, \href
  {http://adsabs.harvard.edu/abs/2016MNRAS.461.4435M} {461, 4435}

\bibitem[\protect\citeauthoryear{{Most}, {Nathanail}  \& {Rezzolla}}{{Most}
  et~al.}{2018}]{Most2017}
{Most} E.~R.,  {Nathanail} A.,   {Rezzolla} L.,  2018, \mn@doi [Astrophys. J]
  {10.3847/1538-4357/aad6ef}, \href
  {http://adsabs.harvard.edu/abs/2018ApJ...864..117M} {864, 117}

\bibitem[\protect\citeauthoryear{{Nathanail}}{{Nathanail}}{2018}]{Nathanail2018}
{Nathanail} A.,  2018, \mn@doi [Astrophys. J.] {10.3847/1538-4357/aad3b8},
  \href {http://adsabs.harvard.edu/abs/2018ApJ...864....4N} {864, 4}

\bibitem[\protect\citeauthoryear{{Nathanail}, {Most}  \&
  {Rezzolla}}{{Nathanail} et~al.}{2017}]{Nathanail2017}
{Nathanail} A.,  {Most} E.~R.,   {Rezzolla} L.,  2017, \mn@doi [Mon. Not. R.
  Astron. Soc.] {10.1093/mnrasl/slx035}, \href
  {http://adsabs.harvard.edu/abs/2017MNRAS.469L..31N} {469, L31}

\bibitem[\protect\citeauthoryear{{Nathanail}, {Porth}  \&
  {Rezzolla}}{{Nathanail} et~al.}{2019}]{Nathanail2018c}
{Nathanail} A.,  {Porth} O.,   {Rezzolla} L.,  2019, \mn@doi [Astrophys. J.
  Lett] {10.3847/2041-8213/aaf73a}, \href
  {https://ui.adsabs.harvard.edu/abs/2019ApJ...870L..20N} {870, L20}

\bibitem[\protect\citeauthoryear{{Palenzuela}}{{Palenzuela}}{2013}]{Palenzuela2013}
{Palenzuela} C.,  2013, \mn@doi [Mon. Not. R. Astron. Soc.]
  {10.1093/mnras/stt311}, \href
  {http://adsabs.harvard.edu/abs/2013MNRAS.431.1853P} {431, 1853}

\bibitem[\protect\citeauthoryear{{Palenzuela}, {Lehner}, {Liebling}, {Ponce},
  {Anderson}, {Neilsen}  \& {Motl}}{{Palenzuela}
  et~al.}{2013a}]{Palenzuela2013b}
{Palenzuela} C.,  {Lehner} L.,  {Liebling} S.~L.,  {Ponce} M.,  {Anderson} M.,
  {Neilsen} D.,   {Motl} P.,  2013a, \mn@doi [Phys. Rev. D]
  {10.1103/PhysRevD.88.043011}, \href
  {http://adsabs.harvard.edu/abs/2013PhRvD..88d3011P} {88, 043011}

\bibitem[\protect\citeauthoryear{{Palenzuela}, {Lehner}, {Ponce}, {Liebling},
  {Anderson}, {Neilsen}  \& {Motl}}{{Palenzuela}
  et~al.}{2013b}]{Palenzuela2013a}
{Palenzuela} C.,  {Lehner} L.,  {Ponce} M.,  {Liebling} S.~L.,  {Anderson} M.,
  {Neilsen} D.,   {Motl} P.,  2013b, \mn@doi [Phys. Rev. Lett.]
  {10.1103/PhysRevLett.111.061105}, \href
  {http://adsabs.harvard.edu/abs/2013PhRvL.111f1105P} {111, 061105}

\bibitem[\protect\citeauthoryear{{Palenzuela}, {Liebling}, {Neilsen}, {Lehner},
  {Caballero}, {O'Connor}  \& {Anderson}}{{Palenzuela}
  et~al.}{2015}]{Palenzuela2015}
{Palenzuela} C.,  {Liebling} S.~L.,  {Neilsen} D.,  {Lehner} L.,  {Caballero}
  O.~L.,  {O'Connor} E.,   {Anderson} M.,  2015, \mn@doi [Phys. Rev. D]
  {10.1103/PhysRevD.92.044045}, \href
  {http://adsabs.harvard.edu/abs/2015PhRvD..92d4045P} {92, 044045}

\bibitem[\protect\citeauthoryear{{Papenfort}, {Gold}  \&
  {Rezzolla}}{{Papenfort} et~al.}{2018}]{Papenfort2018}
{Papenfort} L.~J.,  {Gold} R.,   {Rezzolla} L.,  2018, \mn@doi [Phys. Rev. D]
  {10.1103/PhysRevD.98.104028}, \href
  {https://ui.adsabs.harvard.edu/abs/2018PhRvD..98j4028P} {98, 104028}

\bibitem[\protect\citeauthoryear{Pareschi \& Russo}{Pareschi \&
  Russo}{2005}]{pareschi_2005_ier}
Pareschi L.,  Russo G.,  2005, \mn@doi [Journal of Scientific Computing]
  {10.1007/BF02728986}, 25, 129

\bibitem[\protect\citeauthoryear{{Paschalidis} \& {Ruiz}}{{Paschalidis} \&
  {Ruiz}}{2018}]{Paschalidis2018}
{Paschalidis} V.,  {Ruiz} M.,  2018, preprint, \href
  {http://adsabs.harvard.edu/abs/2018arXiv180804822P} {} (\mn@eprint {arXiv}
  {1808.04822})

\bibitem[\protect\citeauthoryear{{Paschalidis}, {Etienne}, {Liu}  \&
  {Shapiro}}{{Paschalidis} et~al.}{2011}]{Paschalidis2011}
{Paschalidis} V.,  {Etienne} Z.,  {Liu} Y.~T.,   {Shapiro} S.~L.,  2011,
  \mn@doi [Phys. Rev. D] {10.1103/PhysRevD.83.064002}, \href
  {http://adsabs.harvard.edu/abs/2011PhRvD..83f4002P} {83, 064002}

\bibitem[\protect\citeauthoryear{{Philippov} \& {Spitkovsky}}{{Philippov} \&
  {Spitkovsky}}{2014}]{Philippov2014}
{Philippov} A.~A.,  {Spitkovsky} A.,  2014, \mn@doi [Astrophys. J.]
  {10.1088/2041-8205/785/2/L33}, \href
  {https://ui.adsabs.harvard.edu/abs/2014ApJ...785L..33P} {785, L33}

\bibitem[\protect\citeauthoryear{{Piro}}{{Piro}}{2012}]{Piro2012}
{Piro} A.~L.,  2012, \mn@doi [Astrophys. J.] {10.1088/0004-637X/755/1/80},
  \href {http://adsabs.harvard.edu/abs/2012ApJ...755...80P} {755, 80}

\bibitem[\protect\citeauthoryear{{Piro}, {Giacomazzo}  \& {Perna}}{{Piro}
  et~al.}{2017}]{Piro2017}
{Piro} A.~L.,  {Giacomazzo} B.,   {Perna} R.,  2017, \mn@doi [Astrophys. J.
  Lett.] {10.3847/2041-8213/aa7f2f}, \href
  {http://adsabs.harvard.edu/abs/2017ApJ...844L..19P} {844, L19}

\bibitem[\protect\citeauthoryear{{Ponce}, {Palenzuela}, {Lehner}  \&
  {Liebling}}{{Ponce} et~al.}{2014}]{Ponce2014}
{Ponce} M.,  {Palenzuela} C.,  {Lehner} L.,   {Liebling} S.~L.,  2014, \mn@doi
  [Phys. Rev. D] {10.1103/PhysRevD.90.044007}, \href
  {http://adsabs.harvard.edu/abs/2014PhRvD..90d4007P} {90, 044007}

\bibitem[\protect\citeauthoryear{{Pozanenko}, {Minaev}, {Grebenev}  \&
  {Chelovekov}}{{Pozanenko} et~al.}{2019}]{Pozarenko2020}
{Pozanenko} A.~S.,  {Minaev} P.~Y.,  {Grebenev} S.~A.,   {Chelovekov} I.~V.,
  2019, arXiv e-prints, \href
  {https://ui.adsabs.harvard.edu/abs/2019arXiv191213112P} {p. arXiv:1912.13112}

\bibitem[\protect\citeauthoryear{{Radice}, {Galeazzi}, {Lippuner}, {Roberts},
  {Ott}  \& {Rezzolla}}{{Radice} et~al.}{2016}]{Radice2016}
{Radice} D.,  {Galeazzi} F.,  {Lippuner} J.,  {Roberts} L.~F.,  {Ott} C.~D.,
  {Rezzolla} L.,  2016, \mn@doi [Mon. Not. R. Astron. Soc.]
  {10.1093/mnras/stw1227}, \href
  {http://adsabs.harvard.edu/abs/2016MNRAS.tmp..894R} {460, 3255}

\bibitem[\protect\citeauthoryear{{Radice}, {Perego}, {Hotokezaka}, {Fromm},
  {Bernuzzi}  \& {Roberts}}{{Radice} et~al.}{2018}]{Radice2018a}
{Radice} D.,  {Perego} A.,  {Hotokezaka} K.,  {Fromm} S.~A.,  {Bernuzzi} S.,
  {Roberts} L.~F.,  2018, \mn@doi [Astrophys. J.] {10.3847/1538-4357/aaf054},
  \href {https://ui.adsabs.harvard.edu/\#abs/2018ApJ...869..130R} {869, 130}

\bibitem[\protect\citeauthoryear{{Rane} \& {Lorimer}}{{Rane} \&
  {Lorimer}}{2017}]{RaneLorimer2017}
{Rane} A.,  {Lorimer} D.,  2017, \mn@doi [Journal of Astrophysics and
  Astronomy] {10.1007/s12036-017-9478-1}, \href
  {http://adsabs.harvard.edu/abs/2017JApA...38...55R} {38, 55}

\bibitem[\protect\citeauthoryear{{Reisswig}, {Haas}, {Ott}, {Abdikamalov},
  {M{\"o}sta}, {Pollney}  \& {Schnetter}}{{Reisswig}
  et~al.}{2013}]{Reisswig2012b}
{Reisswig} C.,  {Haas} R.,  {Ott} C.~D.,  {Abdikamalov} E.,  {M{\"o}sta} P.,
  {Pollney} D.,   {Schnetter} E.,  2013, \mn@doi [Phys. Rev. D]
  {10.1103/PhysRevD.87.064023}, \href
  {http://adsabs.harvard.edu/abs/2013PhRvD..87f4023R} {87, 064023}

\bibitem[\protect\citeauthoryear{{Rezzolla} \& {Takami}}{{Rezzolla} \&
  {Takami}}{2013}]{Rezzolla2013}
{Rezzolla} L.,  {Takami} K.,  2013, \mn@doi [Class. Quantum Grav.]
  {10.1088/0264-9381/30/1/012001}, \href
  {http://adsabs.harvard.edu/abs/2013CQGra..30a2001R} {30, 012001}

\bibitem[\protect\citeauthoryear{{Rezzolla}, {Giacomazzo}, {Baiotti}, {Granot},
  {Kouveliotou}  \& {Aloy}}{{Rezzolla} et~al.}{2011}]{Rezzolla:2011}
{Rezzolla} L.,  {Giacomazzo} B.,  {Baiotti} L.,  {Granot} J.,  {Kouveliotou}
  C.,   {Aloy} M.~A.,  2011, \mn@doi [Astrophys. J. Letters]
  {10.1088/2041-8205/732/1/L6}, \href
  {http://adsabs.harvard.edu/abs/2011ApJ...732L...6R} {732, L6}

\bibitem[\protect\citeauthoryear{{Ruderman} \& {Sutherland}}{{Ruderman} \&
  {Sutherland}}{1975}]{Ruderman1975}
{Ruderman} M.~A.,  {Sutherland} P.~G.,  1975, \mn@doi [Astrophys. J.]
  {10.1086/153393}, \href {http://adsabs.harvard.edu/abs/1975ApJ...196...51R}
  {196, 51}

\bibitem[\protect\citeauthoryear{{Ruiz} \& {Shapiro}}{{Ruiz} \&
  {Shapiro}}{2017}]{Ruiz2017a}
{Ruiz} M.,  {Shapiro} S.~L.,  2017, \mn@doi [Phys. Rev. D]
  {10.1103/PhysRevD.96.084063}, \href
  {http://adsabs.harvard.edu/abs/2017PhRvD..96h4063R} {96, 084063}

\bibitem[\protect\citeauthoryear{{Ruiz}, {Lang}, {Paschalidis}  \&
  {Shapiro}}{{Ruiz} et~al.}{2016}]{Ruiz2016}
{Ruiz} M.,  {Lang} R.~N.,  {Paschalidis} V.,   {Shapiro} S.~L.,  2016, \mn@doi
  [Astrophys. J. Lett.] {10.3847/2041-8205/824/1/L6}, \href
  {http://adsabs.harvard.edu/abs/2016ApJ...824L...6R} {824, L6}

\bibitem[\protect\citeauthoryear{{Ruiz}, {Shapiro}  \& {Tsokaros}}{{Ruiz}
  et~al.}{2018}]{Ruiz2017}
{Ruiz} M.,  {Shapiro} S.~L.,   {Tsokaros} A.,  2018, \mn@doi [Phys. Rev. D]
  {10.1103/PhysRevD.97.021501}, \href
  {http://adsabs.harvard.edu/abs/2018PhRvD..97b1501R} {97, 021501}

\bibitem[\protect\citeauthoryear{{Schnetter}, {Hawley}  \& {Hawke}}{{Schnetter}
  et~al.}{2004}]{Schnetter-etal-03b}
{Schnetter} E.,  {Hawley} S.~H.,   {Hawke} I.,  2004, \mn@doi [Class. Quantum
  Grav.] {10.1088/0264-9381/21/6/014}, \href
  {http://adsabs.harvard.edu/abs/2004CQGra..21.1465S} {21, 1465}

\bibitem[\protect\citeauthoryear{{Sekiguchi}, {Kiuchi}, {Kyutoku}  \&
  {Shibata}}{{Sekiguchi} et~al.}{2015}]{Sekiguchi2015}
{Sekiguchi} Y.,  {Kiuchi} K.,  {Kyutoku} K.,   {Shibata} M.,  2015, \mn@doi
  [Phys. Rev. D] {10.1103/PhysRevD.91.064059}, \href
  {http://adsabs.harvard.edu/abs/2015PhRvD..91f4059S} {91, 064059}

\bibitem[\protect\citeauthoryear{{Sekiguchi}, {Kiuchi}, {Kyutoku}, {Shibata}
  \& {Taniguchi}}{{Sekiguchi} et~al.}{2016}]{Sekiguchi2016}
{Sekiguchi} Y.,  {Kiuchi} K.,  {Kyutoku} K.,  {Shibata} M.,   {Taniguchi} K.,
  2016, \mn@doi [Phys. Rev. D] {10.1103/PhysRevD.93.124046}, \href
  {http://adsabs.harvard.edu/abs/2016PhRvD..93l4046S} {93, 124046}

\bibitem[\protect\citeauthoryear{{Shibata} \& {Taniguchi}}{{Shibata} \&
  {Taniguchi}}{2006}]{Shibata06a}
{Shibata} M.,  {Taniguchi} K.,  2006, \mn@doi [Phys. Rev. D]
  {10.1103/PhysRevD.73.064027}, \href
  {http://adsabs.harvard.edu/abs/2006PhRvD..73f4027S} {73, 064027}

\bibitem[\protect\citeauthoryear{{Shibata}, {Suwa}, {Kiuchi}  \&
  {Ioka}}{{Shibata} et~al.}{2011}]{Shibata2011b}
{Shibata} M.,  {Suwa} Y.,  {Kiuchi} K.,   {Ioka} K.,  2011, \mn@doi [Astrophys.
  J.l] {10.1088/2041-8205/734/2/L36}, \href
  {http://adsabs.harvard.edu/abs/2011ApJ...734L..36S} {734, L36}

\bibitem[\protect\citeauthoryear{{Siegel} \& {Metzger}}{{Siegel} \&
  {Metzger}}{2018}]{Siegel2018}
{Siegel} D.~M.,  {Metzger} B.~D.,  2018, \mn@doi [Astrophys. J.]
  {10.3847/1538-4357/aabaec}, \href
  {http://adsabs.harvard.edu/abs/2018ApJ...858...52S} {858, 52}

\bibitem[\protect\citeauthoryear{{Sturrock}}{{Sturrock}}{1971}]{Sturrock1971}
{Sturrock} P.~A.,  1971, \mn@doi [Astrophys. J.] {10.1086/150865}, \href
  {http://adsabs.harvard.edu/abs/1971ApJ...164..529S} {164, 529}

\bibitem[\protect\citeauthoryear{{Szary}, {Zhang}, {Melikidze}, {Gil}  \&
  {Xu}}{{Szary} et~al.}{2014}]{Szary2014}
{Szary} A.,  {Zhang} B.,  {Melikidze} G.~I.,  {Gil} J.,   {Xu} R.-X.,  2014,
  \mn@doi [Astrophys. J.] {10.1088/0004-637X/784/1/59}, \href
  {https://ui.adsabs.harvard.edu/abs/2014ApJ...784...59S} {784, 59}

\bibitem[\protect\citeauthoryear{{The LIGO Scientific Collaboration}
  et~al.,}{{The LIGO Scientific Collaboration} et~al.}{2020}]{Abbott2020}
{The LIGO Scientific Collaboration} et~al., 2020, arXiv e-prints, \href
  {https://ui.adsabs.harvard.edu/abs/2020arXiv200101761T} {p. arXiv:2001.01761}

\bibitem[\protect\citeauthoryear{{Wang}, {Yang}, {Wu}, {Dai}  \& {Wang}}{{Wang}
  et~al.}{2016}]{Wang2016}
{Wang} J.-S.,  {Yang} Y.-P.,  {Wu} X.-F.,  {Dai} Z.-G.,   {Wang} F.-Y.,  2016,
  \mn@doi [Astrophys. J. Lett.] {10.3847/2041-8205/822/1/L7}, \href
  {http://adsabs.harvard.edu/abs/2016ApJ...822L...7W} {822, L7}

\bibitem[\protect\citeauthoryear{{Zhang}}{{Zhang}}{2016}]{Zhang2016}
{Zhang} B.,  2016, \mn@doi [Astrophys. J. Lett.] {10.3847/2041-8205/827/2/L31},
  \href {http://adsabs.harvard.edu/abs/2016ApJ...827L..31Z} {827, L31}

\makeatother
\end{thebibliography}






\end{document}